\documentclass[12pt,preprint]{aastex}

\shorttitle{Gas Absorption in the KH~15D System}
\shortauthors{Lawler et al.}

\begin{document}

\title{Gas Absorption in the KH~15D System: Further Evidence for Dust Settling in the Circumbinary Disk}

\author{S.~M.~Lawler\altaffilmark{1,7}, W.~Herbst\altaffilmark{1}, S.~Redfield\altaffilmark{1}, C.~M.~Hamilton\altaffilmark{2}, C.~M.~Johns-Krull\altaffilmark{3}, J.~N.~Winn\altaffilmark{4}, J.~A.~Johnson\altaffilmark{5}, R.~Mundt\altaffilmark{6}}
\altaffiltext{1}{Astronomy Department, Wesleyan University, Middletown, CT 06459 USA}
\altaffiltext{2}{Physics and Astronomy Department, Dickinson College, Carlisle, PA 17013 USA}
\altaffiltext{3}{Department of Physics and Astronomy, Rice University, 6100 Main Street, Houston, TX 77005 USA}
\altaffiltext{4}{Department of Physics and Kavli Institute for Astrophysics and Space Research, Massachusetts Institute of Technology, Cambridge, MA 02139 USA}
\altaffiltext{5}{Department of Astrophysics, California Institute of Technology, MC 249-17, Pasadena, CA 91125 USA}
\altaffiltext{6}{Max-Planck-Institute f{\"u}r Astronomie, K{\"o}nigstuhl 17, D-69117 Heidelberg, Germany}
\altaffiltext{7}{Current address: Department of Physics and Astronomy, University of British Columbia, 6244 Agricultural Road, Vancouver, BC V6T 1Z1 Canada}

\begin{abstract}
\ion{Na}{1}~D lines in the spectrum of the young binary KH~15D have been analyzed in detail. We find an excess absorption component that may be attributed to foreground interstellar absorption, and to gas possibly associated with the solids in the circumbinary disk. The derived column density is log~$N_{\rm NaI}$ = 12.5~cm$^{-2}$, centered on a radial velocity that is consistent with the systemic velocity.  Subtracting the likely contribution of the ISM leaves log~$N_{\rm NaI} \sim$ 12.3~cm$^{-2}$. There is no detectable change in the gas column density across the ``knife edge" formed by the opaque grain disk, indicating that the gas and solids have very different scale heights, with the solids being highly settled. Our data support a picture of this circumbinary disk as being composed of a very thin particulate grain layer composed of millimeter-sized or larger objects that are settled within whatever remaining gas may be present. This phase of disk evolution has been hypothesized to exist as a prelude to the formation of planetesimals through gravitational fragmentation, and is expected to be short-lived if much gas were still present in such a disk. Our analysis also reveals the presence of excess \ion{Na}{1} emission relative to the comparison spectrum at the radial velocity of the currently visible star that plausibly arises within the magnetosphere of this still-accreting young star.  
\end{abstract}

\section{Introduction}

KH~15D is a system with unique and dramatic photometric behavior that was discovered by \citet{kearns98} during a variability study of the extremely young cluster NGC~2264.  Basic properties of the star, including its distance (760~pc) and classification as a weak-lined T~Tauri star (WTTS) were reported by \citet{hamilton01}. It is now known to be a young ($\sim$3~Myr) eccentric binary system embedded in a nearly edge-on circumbinary disk that is tilted, warped and precessing with respect to the binary orbit \citep{winn04, johnsonwinn04, chiang04, johnson04}.  When first discovered in 1995, the disk was oriented relative to our line of sight in such a way that it covered most of the orbit of one star (now known as Star~B) but only a small portion of the orbit of the other (Star~A). During the past 14 years, precession of the disk has caused the sharp edge of the opaque screen to gradually cover all of the orbit of Star~B and most of the orbit of Star~A. As a result, we have not seen Star~B (except by reflected light) since 1995, and have seen direct light from Star~A for progressively less of its 48.4 day orbital cycle \citep{hamilton05, leduc09}. \citet{winn06} have produced a model of the system that accurately accounts for most of the photometric and radial velocity observations. They estimate the masses of Star~A and Star~B to be 0.6 and 0.7 $M_{\odot}$ and the radii to be 1.3 and 1.4 $R_{\odot}$, respectively. The spectral class of Star~A is K6/K7 \citep{hamilton01, agol04}. 

The system's geometry and fortunate orientation have provided astronomers with a unique opportunity during the past decade and a half to study the structure of both the gas and the dust components of an evolved protoplanetary disk. \citet{herbst08} argued that the particulate grains making up the solid component of the disk, the ``grain disk", are probably of sand (millimeter) size or larger.  It is possible that KH~15D possesses exactly the sort of settled particulate disk originally envisioned by \citet{safronov69} and \citet{goldreich73} as a prelude to the formation of planetesimals through gravitational instability \citep[see also][]{youdin08}. The gas disk at this stage is predicted to have a much larger scale height than the particulate disk since it is partially supported by pressure.  Because Star~A of the KH~15D system changes elevations with respect to the edge of the grain disk, during the past decade we were presented with an opportunity to observe different path lengths through any accompanying gas disk for signs of absorption. We know that there is some gas still present in the disk because the star has an active magnetosphere and drives a weak bipolar jet \citep{hamilton03,tokunaga04,deming04,mundt09,hamilton09}. KH~15D has the potential to tell us about an important missing link in the planet formation process: the transition from protoplanetary disk to debris disk.   

In this paper, we constrain properties of the gas component of the KH~15D circumbinary disk by carefully analyzing the \ion{Na}{1}~D absorption lines visible in the spectrum of Star~A while at various heights above the occulting disk. In \S\ref{observations} we describe the echelle spectra available for this analysis, their reduction, and the basic features of the observed \ion{Na}{1} absorption profile. In \S\ref{measuring} we discuss how \ion{Na}{1} column densities are computed. In \S\ref{ISM} we measure interstellar \ion{Na}{1} absorption toward KH~15D and discuss how this affects our absorption measurements. In \S\ref{scaleheight} we discuss the implications for the scale height of the gas disk, and in \S\ref{graindisk} we further discuss the interpretation of the data in terms of what it may reveal about the putative gas disk of KH~15D. Finally, in \S\ref{summary} we summarize the insight into the KH~15D disk revealed by this investigation. 

\section{Observations} \label{observations}

\subsection{Spectral Data}

Over the past few years, a number of high-resolution echelle spectra of KH~15D have been obtained with Star~A at different elevations with respect to the grain disk edge.  Some of these spectra include the \ion{Na}{1}~D doublet at 5895.9242 and 5889.9510 \AA, which allows us to search for evidence of absorption by neutral sodium gas which may be associated with the disk.  These spectra were taken with the High Resolution Echelle Spectrometer \citep[HIRES;][]{vogt94} on Keck~I at the W.~M.\ Keck Observatory in Hawaii, the Ultraviolet and Visual Echelle Spectrograph \citep[UVES;][]{dekker00} on the Kueyen Telescope at the European Southern Observatory's Very Large Telescope (VLT) facility on Cerro Paranal in Chile, and the High Resolution Spectrograph \citep[HRS;][]{tull98} on the Hobby-Eberly Telescope \citep[HET;][]{ramsey98} at McDonald Observatory in Texas.  A log of the spectra analyzed in this paper is given in Table~\ref{cds}.


The Keck/HIRES data were taken at a resolution $R \equiv \lambda/\Delta\lambda \sim 45,000$ in 2003 February and 2005 February (W.~Herbst, P.I.)  The reader is referred to \citet{hamilton09} for a full desciption of the reduction procedures employed. The VLT/UVES data were taken at a resolution $R \sim 44,000$ in two separate observing runs, one in 2001 November (C.~Bailer-Jones, P.I.) and one in 2004 December (R.~Mundt, P.I.)  Data were reduced using the UVES pipeline as described in \citet{hamilton03} and \citet{hamilton09}.  The HET/HRS data were taken at a resolution $R \sim 30,000$ in 2005 December and 2006 February (W.~Herbst, P.I.)  Data were reduced using IRAF \citep{IRAF} routines to subtract the bias, perform flat-field corrections, remove scattered light and cosmic rays, extract the echelle science and sky orders, and calibrate the wavelength solution using images of a Th-Ar hollow cathode taken the same night as the science data.  IDL routines were written to subtract the sky spectra, normalize the science spectra, and convert to heliocentric velocities.  

The comparison spectrum used here is of HD~36003, which has a spectral type of K5~V \citep{cenarro07}.  The spectrum was taken on the night of 2003 February 8 with Keck/HIRES. This star has a parallax of 0.077\arcsec\ \citep{hipparcos}, making it close enough to alleviate worries about significant ISM contamination. While the spectral type is slightly earlier than that of the visible component of KH~15D (K6/K7), the match with other photospheric lines and with the extreme wings of the \ion{Na}{1}~D lines is excellent. The slightly later spectral class of KH~15D is offset by its lower surface gravity relative to a main sequence K6/K7 star. 

We initially attempted to use other comparison stars that more closely match KH~15D's spectral type and evolutionary phase, but HD~36003 had the best overall fit to the spectra of KH~15D.  Spectra of Hubble~4 and LkCa~7, which are nominally T~Tauri stars (TTS) of spectral class K7, were compared with KH~15D, but the fit with the clearly uncontaminated spectral features was significantly worse for both T~Tauri spectra than for HD~36003.  This is most likely due to the larger rotational broadening of the features in these two TTS as compared to KH~15D, which has $v$~sin~$i$ = 6.9~$\pm$~0.3~km/s \citep{hamilton05}, and HD~36003, which has $v$~sin~$i$ $<$ 3~km/s \citep{luck06}. Hubble~4 has $v$~sin~$i$ = 14.6~$\pm$~1.7~km/s \citep{johnskrull04}, while LkCa~7 has $v$~sin~$i$ = 11~km/s \citep{glebocki01}.  In addition, LkCa~7 has been noted to be a poor comparison star due to anomalous colors, possibly caused by a cooler companion \citep{gullbring98}.  For these reasons, we chose to use HD~36003 as our comparison star.

In addition to this comparison star, the bright, nearby B2~III star HD~47887 was observed to place constraints on interstellar \ion{Na}{1} absorption near KH~15D (\S\ref{ISM}).

\subsection{The \ion{Na}{1}~D Profile of KH~15D} \label{profile}



\begin{figure}
\centering
\includegraphics[scale=0.3]{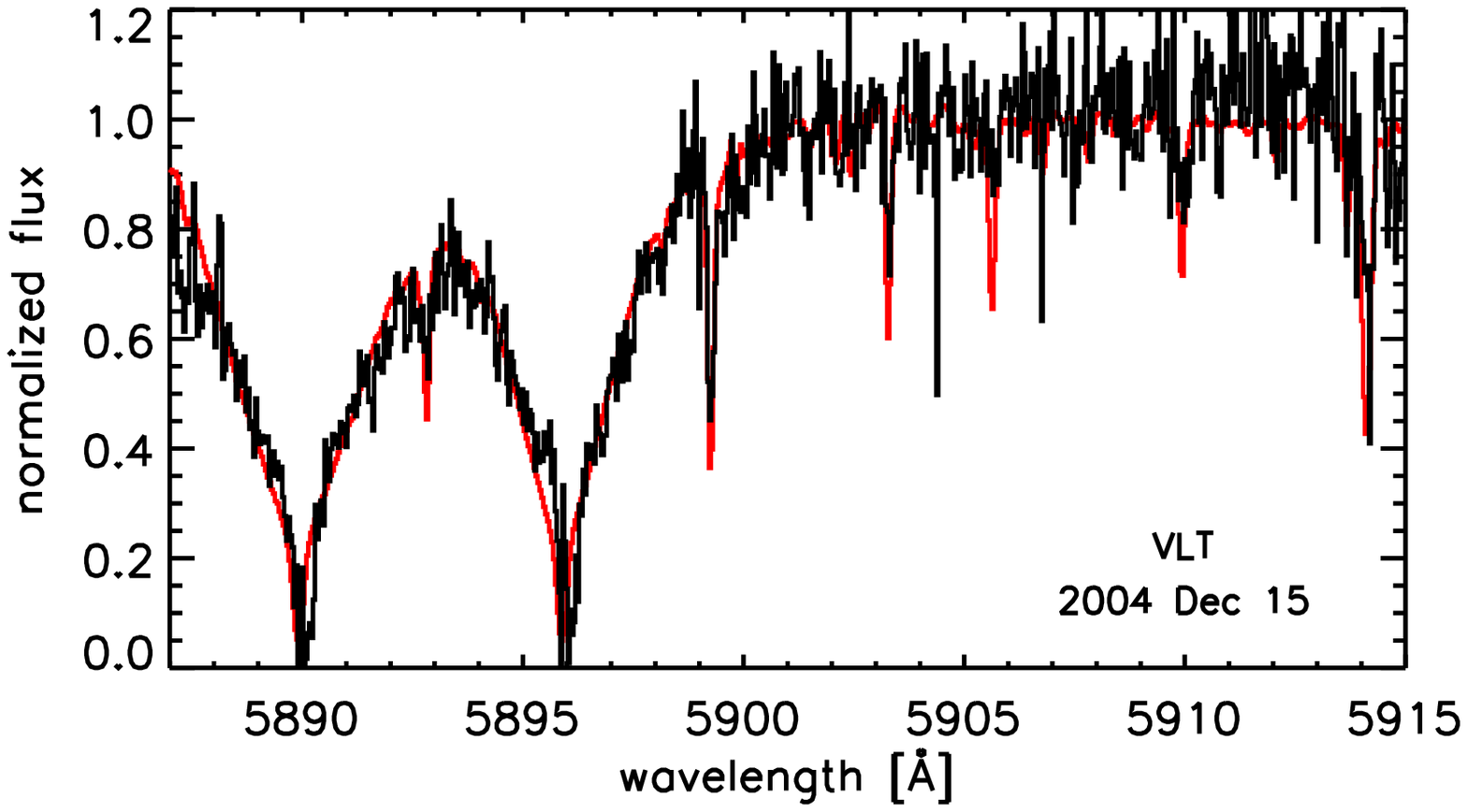}\includegraphics[scale=0.3]{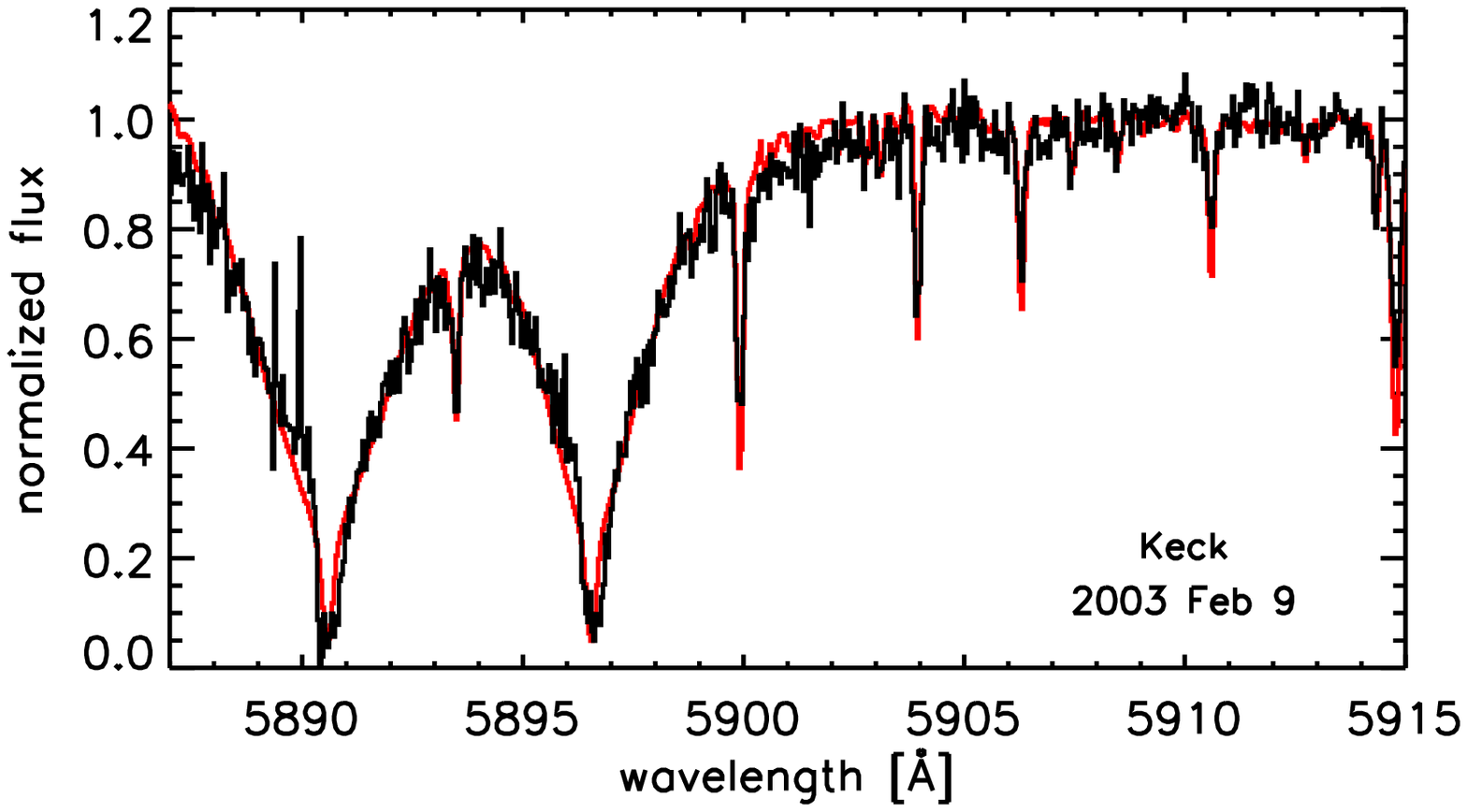}\includegraphics[scale=0.3]{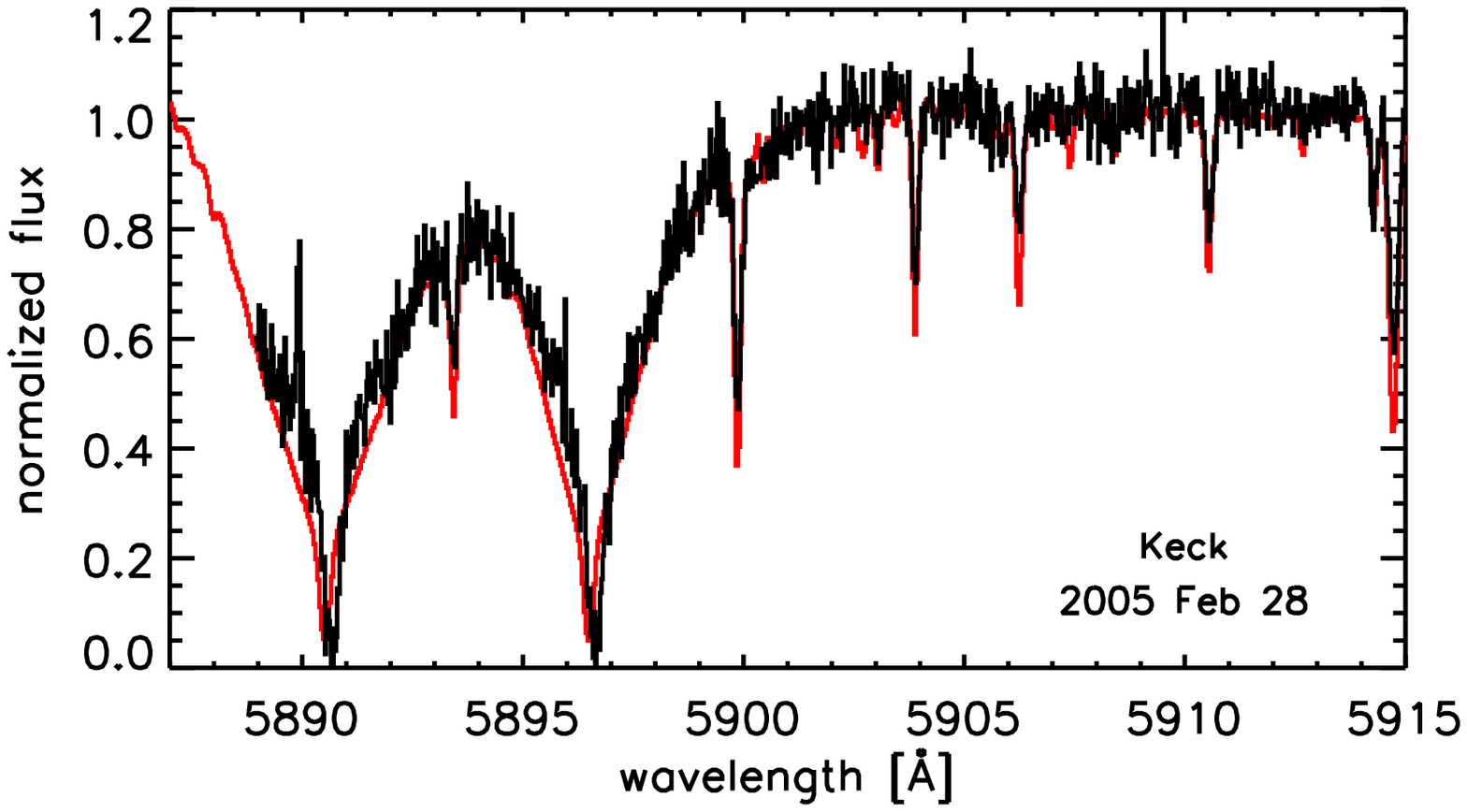} 
\includegraphics[scale=0.3]{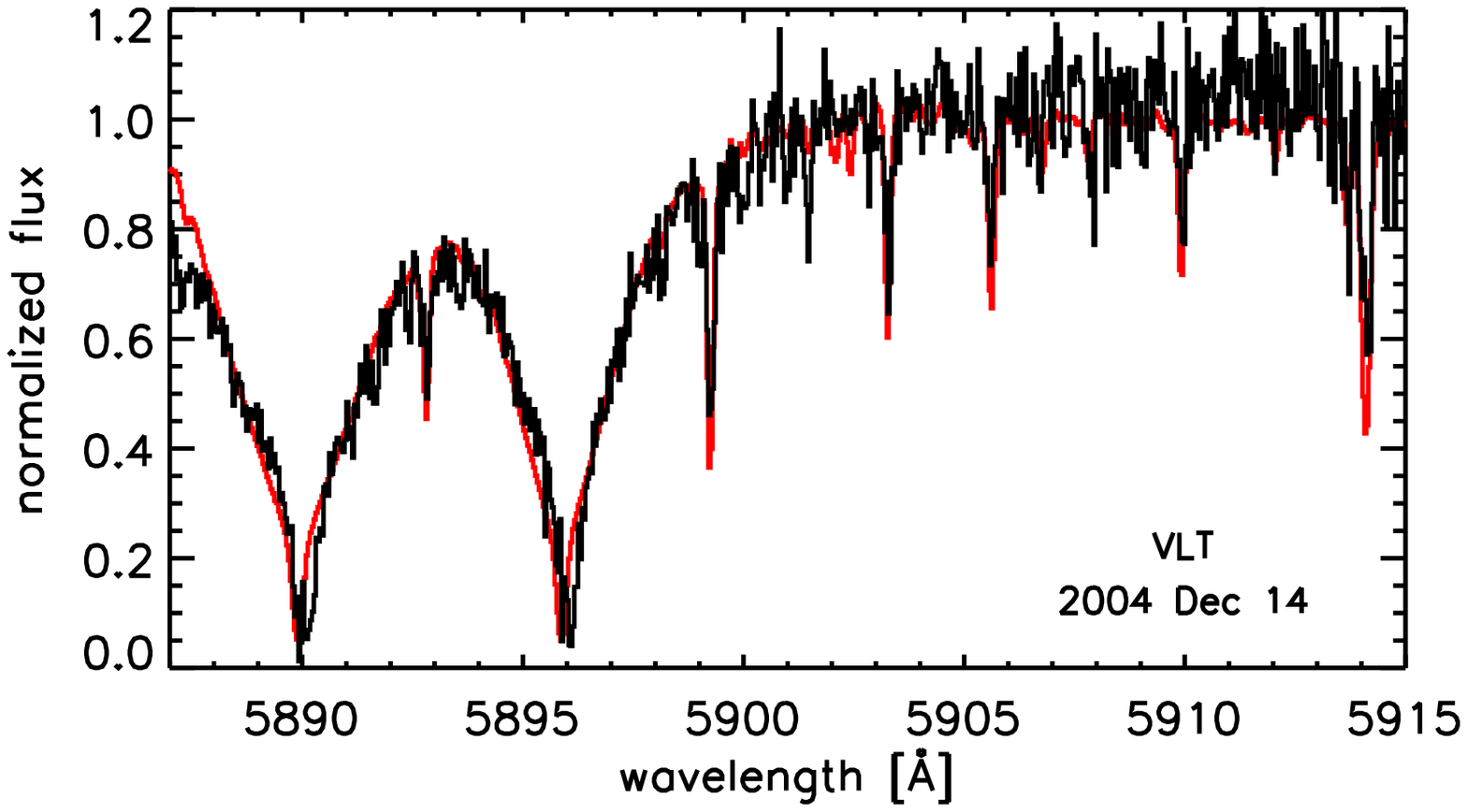}\includegraphics[scale=0.3]{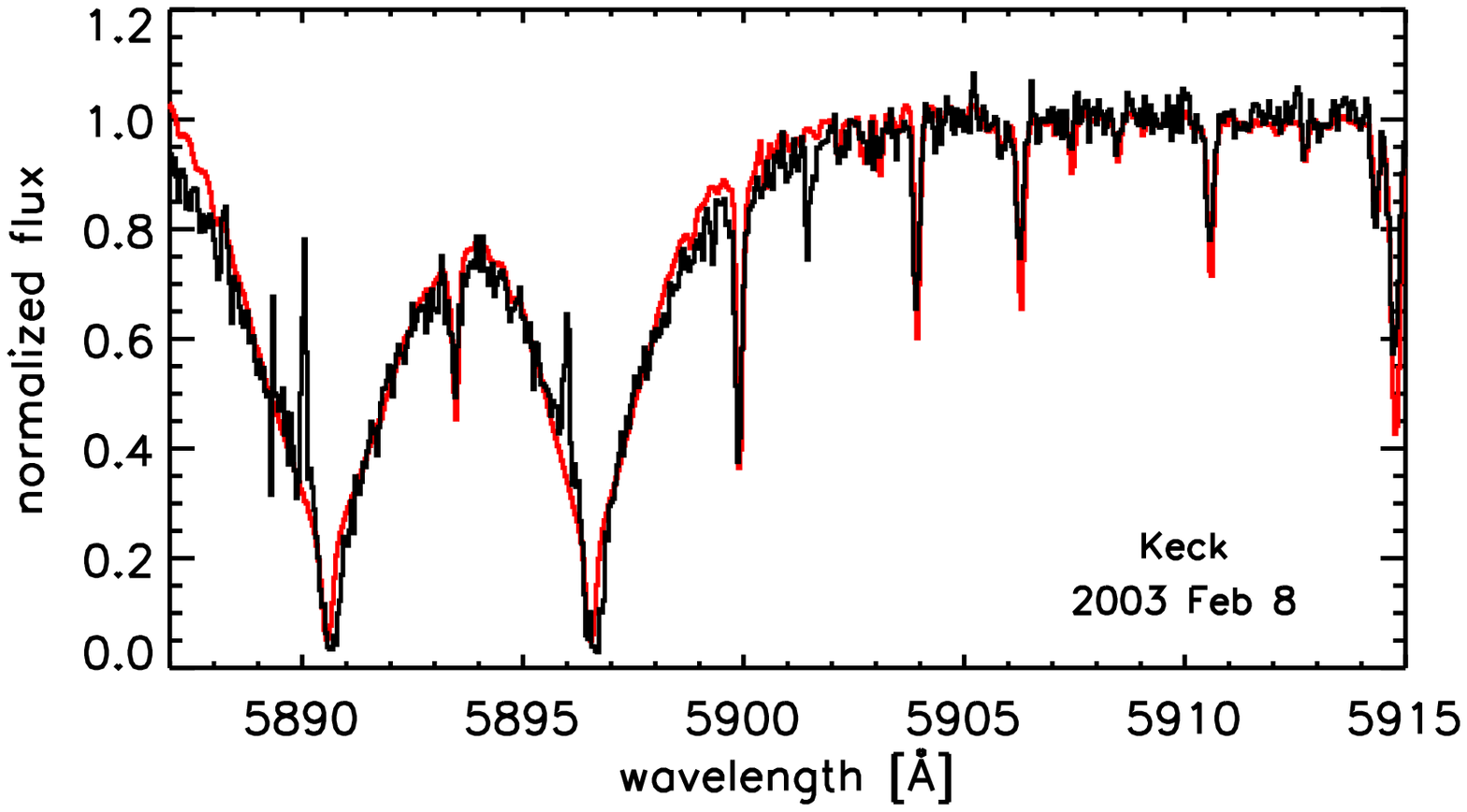}\includegraphics[scale=0.3]{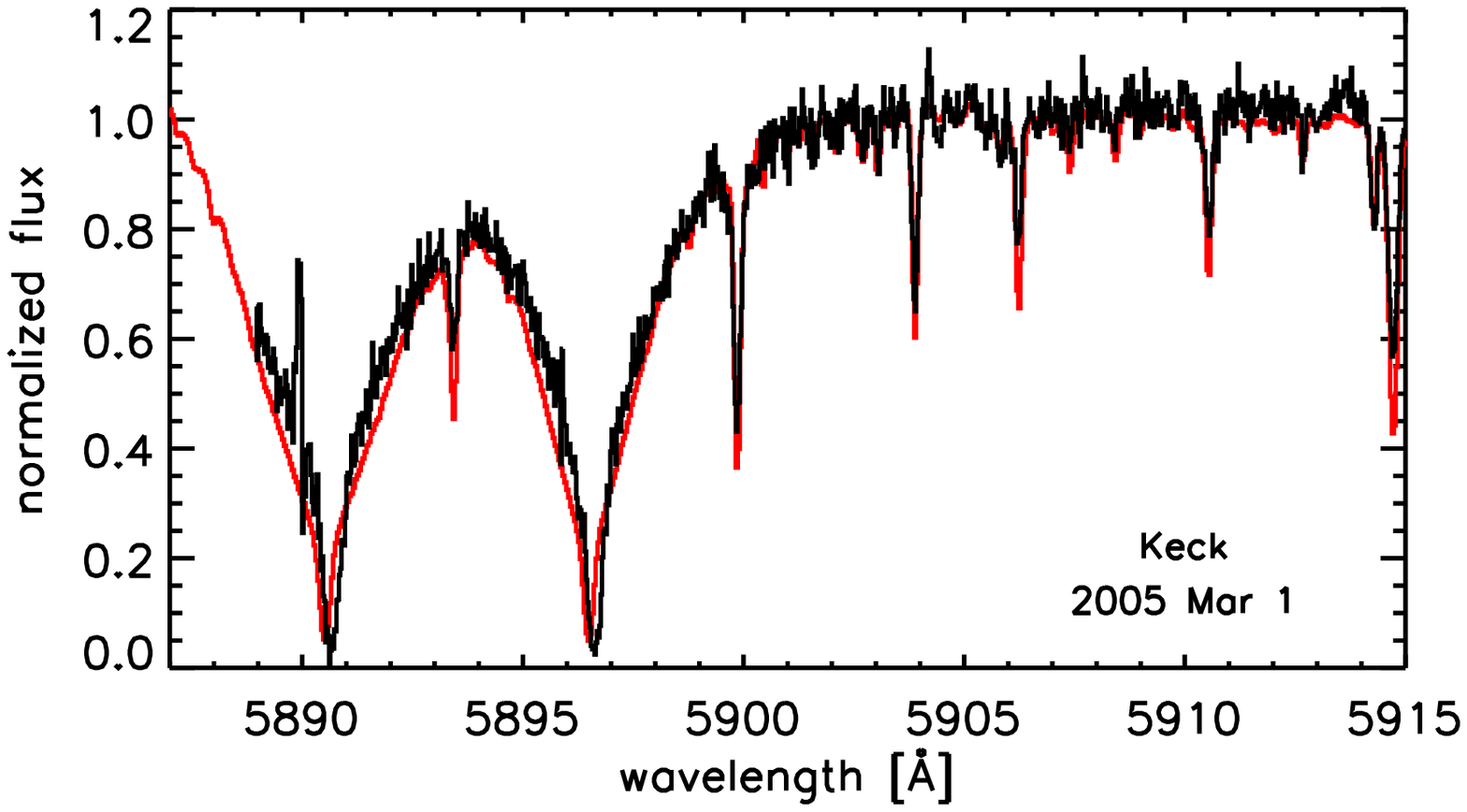} 
\includegraphics[scale=0.3]{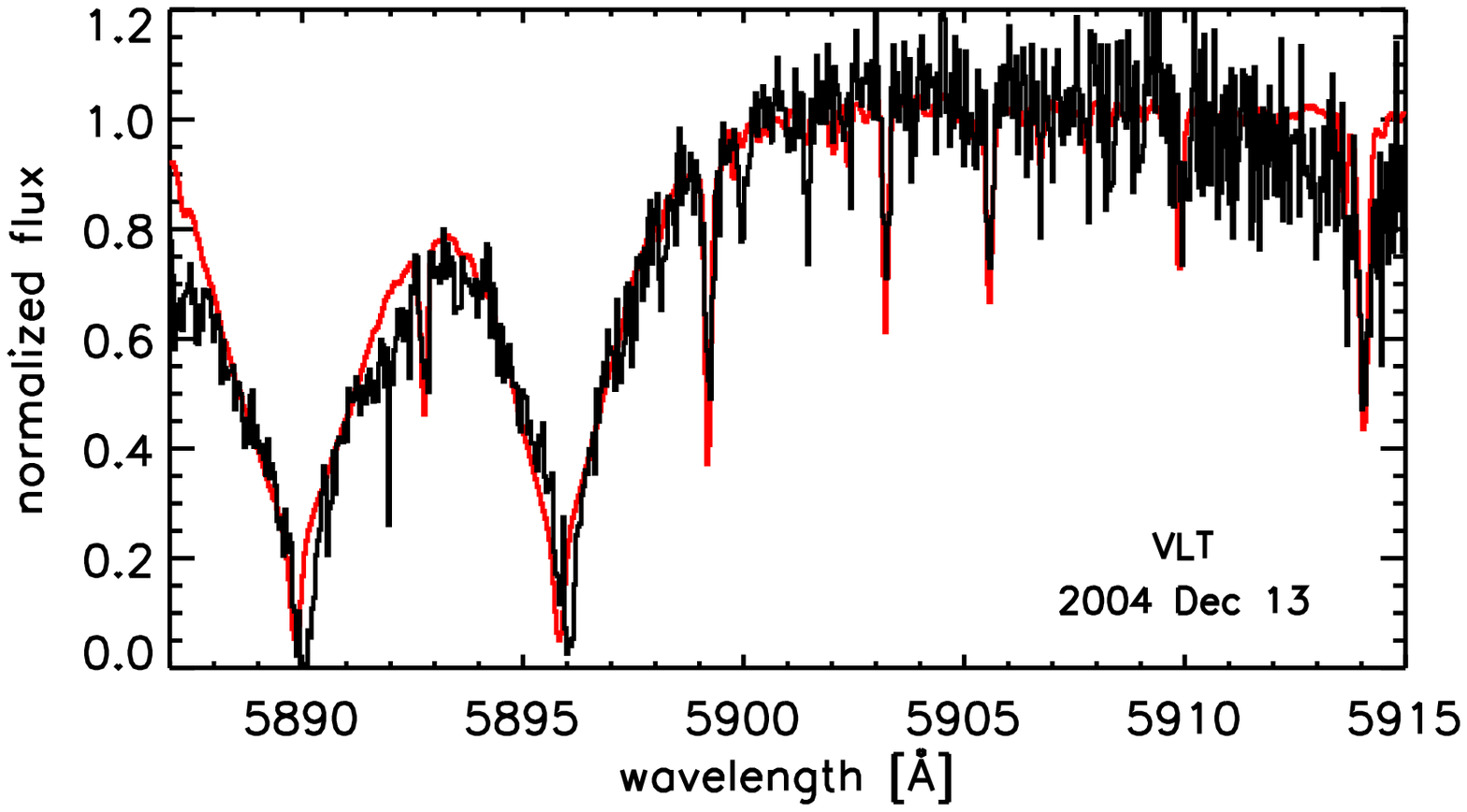}\includegraphics[scale=0.3]{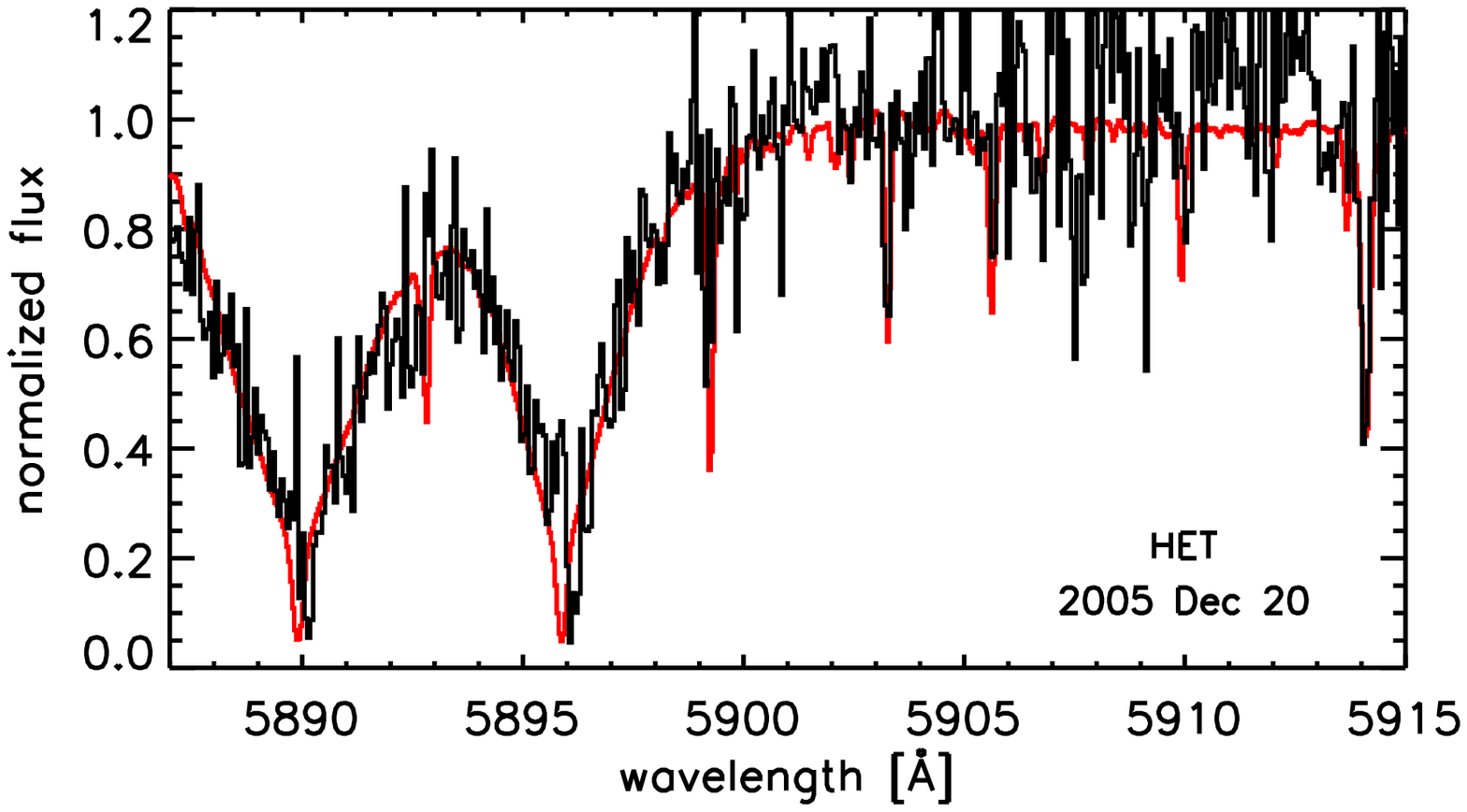}\includegraphics[scale=0.3]{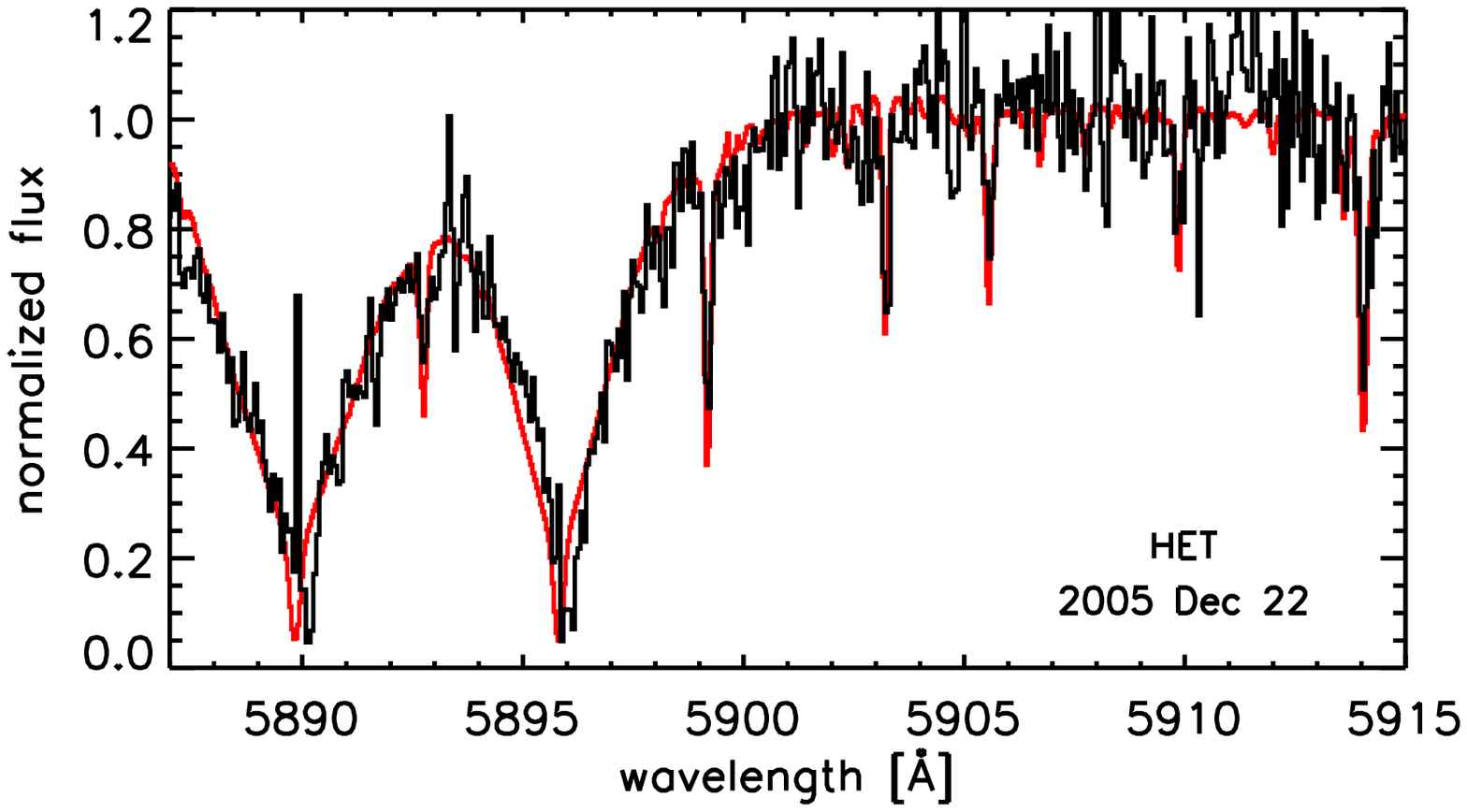} 
\includegraphics[scale=0.3]{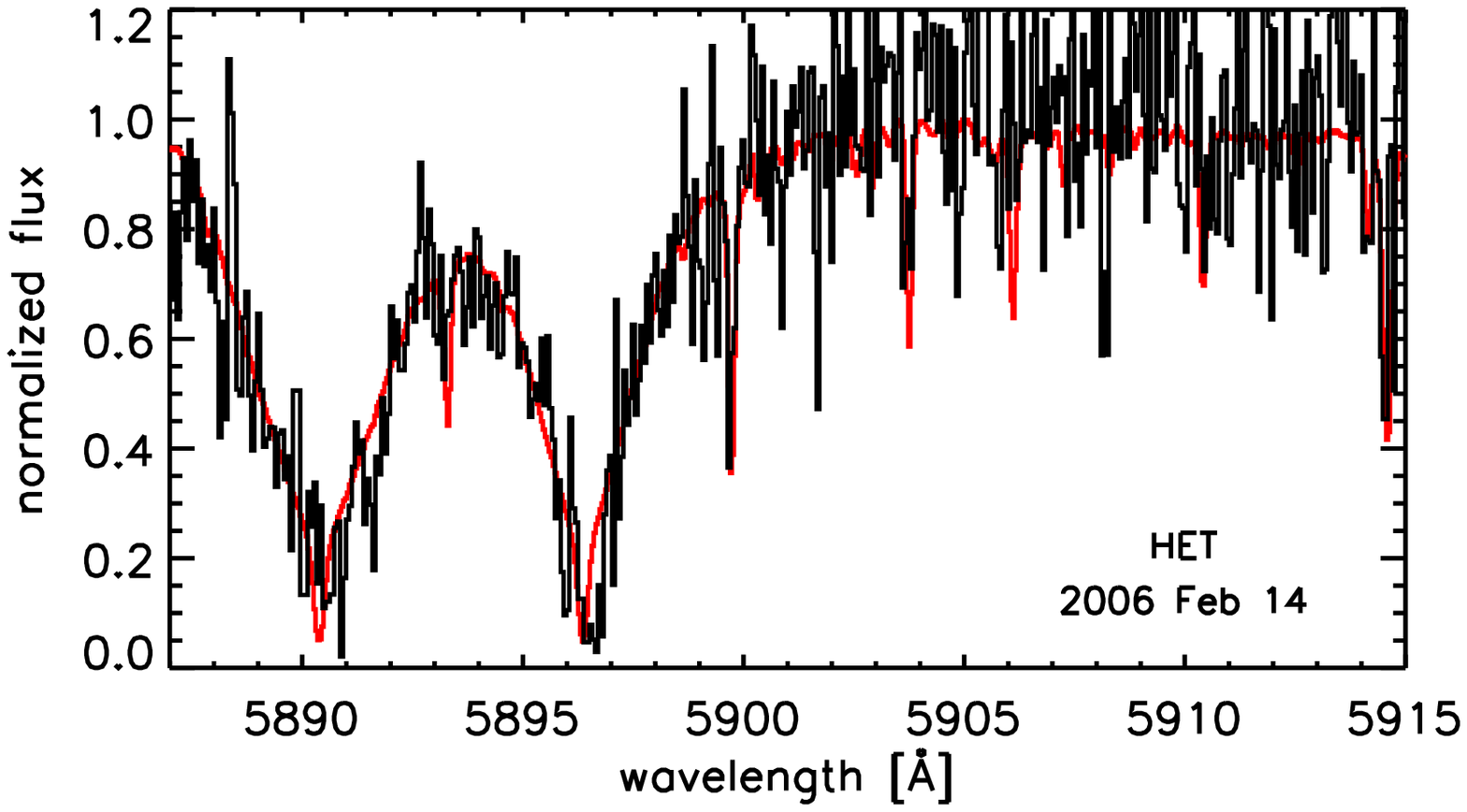}\includegraphics[scale=0.3]{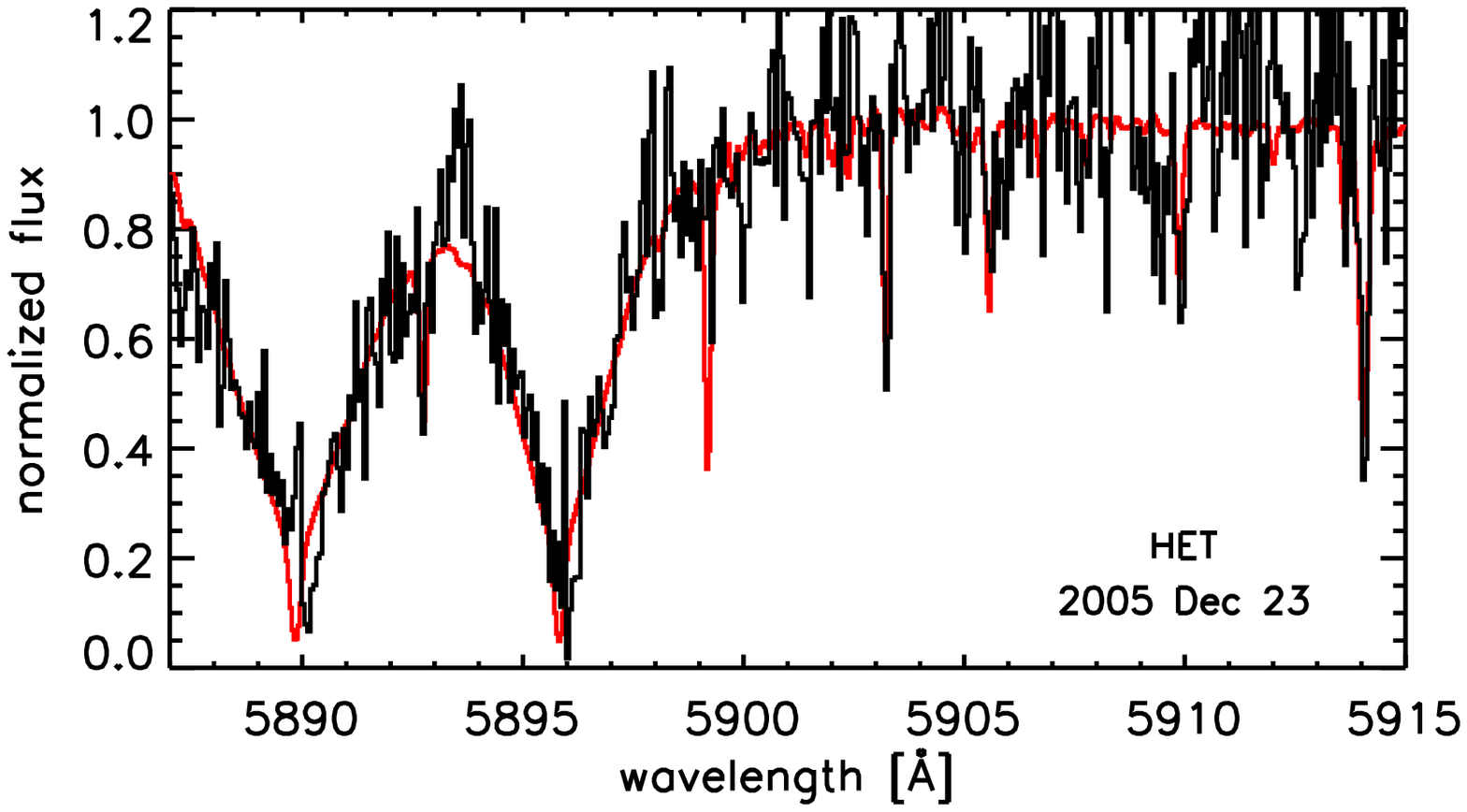}\includegraphics[scale=0.3]{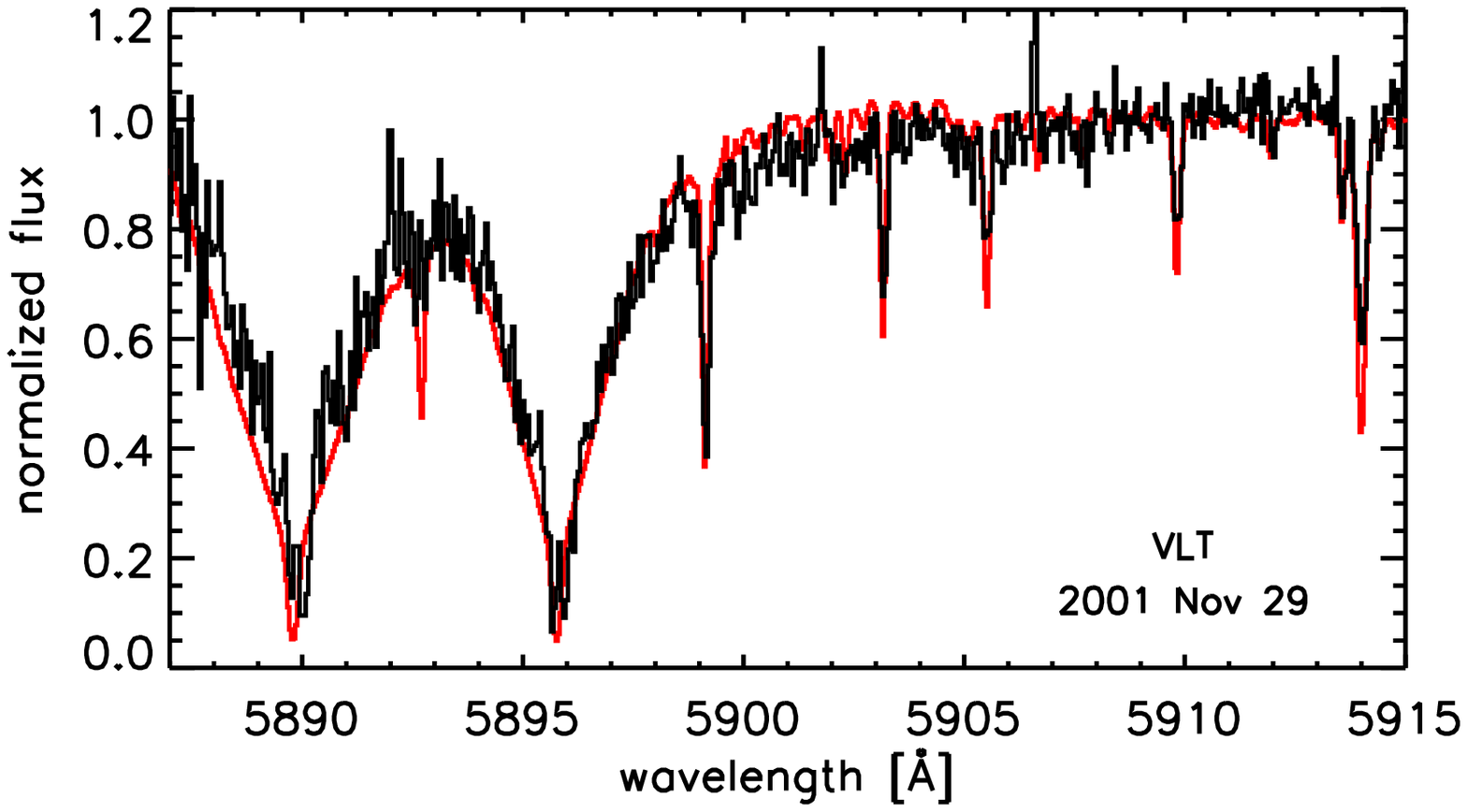} 
\caption{Spectra of KH~15D from different nights and the comparison star spectrum, which has been shifted to match the KH~15D spectra.  The KH~15D spectra are shown in black, while the comparison star spectrum is red.  Nights are arranged in order of increasing $\Delta$ (see Table~\ref{cds}).  Note that absorption features are very well aligned, and that the \ion{Na}{1}~D lines both show obvious excess absorption on the positive radial velocity side.}
\label{compspec_sm}
\end{figure}

Figure~\ref{compspec_sm} shows each of the analyzed KH~15D spectra in the region of the \ion{Na}{1}~D lines overlaid on the comparison spectrum, which has been appropriately shifted for each night of data. To accurately align the comparison star spectrum in radial velocity with the KH~15D spectra,  we used a $\chi^2$ minimization technique based on the portion of the spectrum longward of the \ion{Na}{1}~D$_1$ line ($\sim$5900--5930~\AA), which contains several stellar absorption features due to \ion{Fe}{1} and \ion{Ti}{1}.  Formally, the comparison star spectrum should be shifted by steps in log($\lambda$) \citep{tonry79}, but beause our analysis region includes such a small range of wavelengths ($\sim$15~\AA), the $\chi^2$ analysis is performed using uniform steps in $\lambda$.  The part of the spectrum within the wings of the \ion{Na}{1}~D lines was excluded from the fitting, because we are attempting to measure very small discrepancies in the deepest part of these lines.

\begin{figure}
\centering
\includegraphics[scale=0.8]{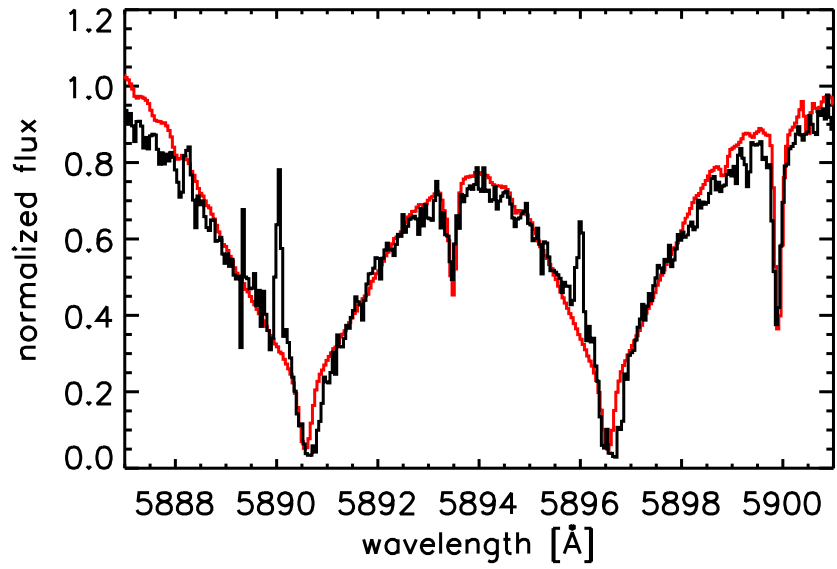} 
\caption{Zoomed in view of the spectrum from Keck taken on 2003 February 8 (black) and the properly shifted comparison star spectrum (red), allowing closer inspection of the absorption and emission features.}
\label{compspec_big}
\end{figure}

Visual inspection of the KH~15D spectra overplotted on the shifted comparison star spectrum clearly reveals that the two match well except at the very bottom of the \ion{Na}{1}~D lines. Figure~\ref{compspec_big} shows an enlarged view of the region around the \ion{Na}{1}~D lines to make the absorption and emission features more clearly visible.  Here we find that the KH~15D \ion{Na}{1}~D line cores deviate from those of the comparison star spectrum, showing some potential filling on the blue side (though telluric emission makes this difficult to see in some panels of Figure~\ref{compspec_sm}) and excess absorption on the red side of the profile. Since this absorption feature appears in all 12 spectra, regardless of which telescope or reduction procedure was employed, we are confident that it is real. As this absorption is asymmetric and not seen in the comparison star spectrum, this additional sodium absorption is most likely caused by something other than the photosphere of Star~A. We turn now to an examination of the source of this absorption and, in particular, whether it could be associated with the gaseous component of the circumbinary disk.

\section{Analysis} \label{analysis}
		
\subsection{Measuring the \ion{Na}{1}~D Profile} \label{measuring}

Following the procedure of \citet{redfield07}, we began the process of quantifying the excess \ion{Na}{1} absorption in KH~15D by taking the ratio between each available spectrum and the shifted comparison star spectrum after the comparison star spectrum had been resampled to have the same pixel sizes as the KH~15D spectrum by rebinning.  The VLT and HET spectra are both at lower resolutions than the Keck spectra, so the comparison spectrum was only resampled to larger pixel sizes. The results of this process are shown in  Figures~\ref{ratiospec_sm} and~\ref{ratio_big} where we have effectively taken out the baseline stellar spectrum, leaving only the excess absorption and emission.  We discuss each of these in turn in \S\S\ref{naemission} and \ref{quant}.



\subsubsection{\ion{Na}{1} Emission}\label{naemission}

\begin{figure}
\centering
\includegraphics[scale=0.4]{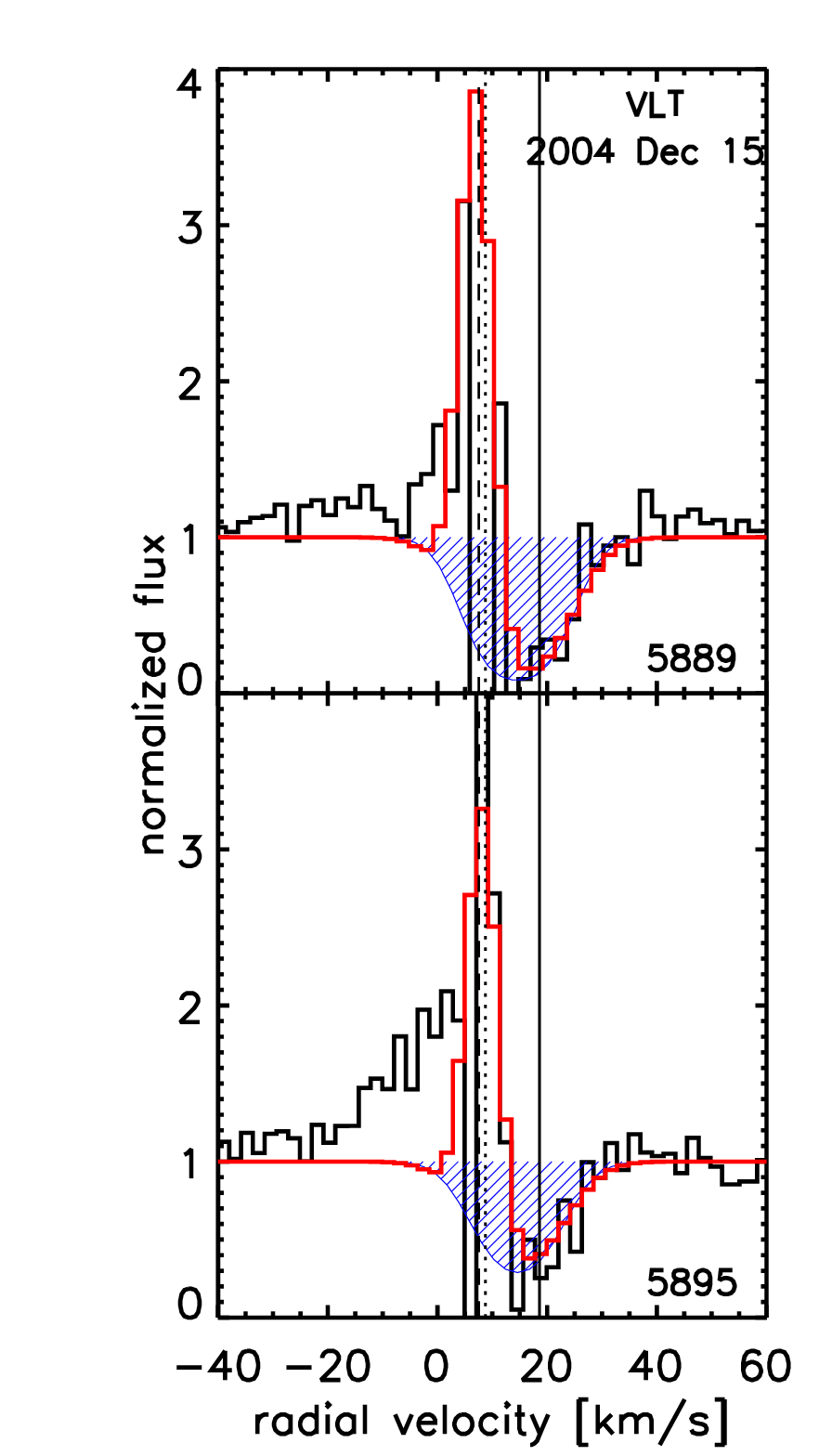}\includegraphics[scale=0.4]{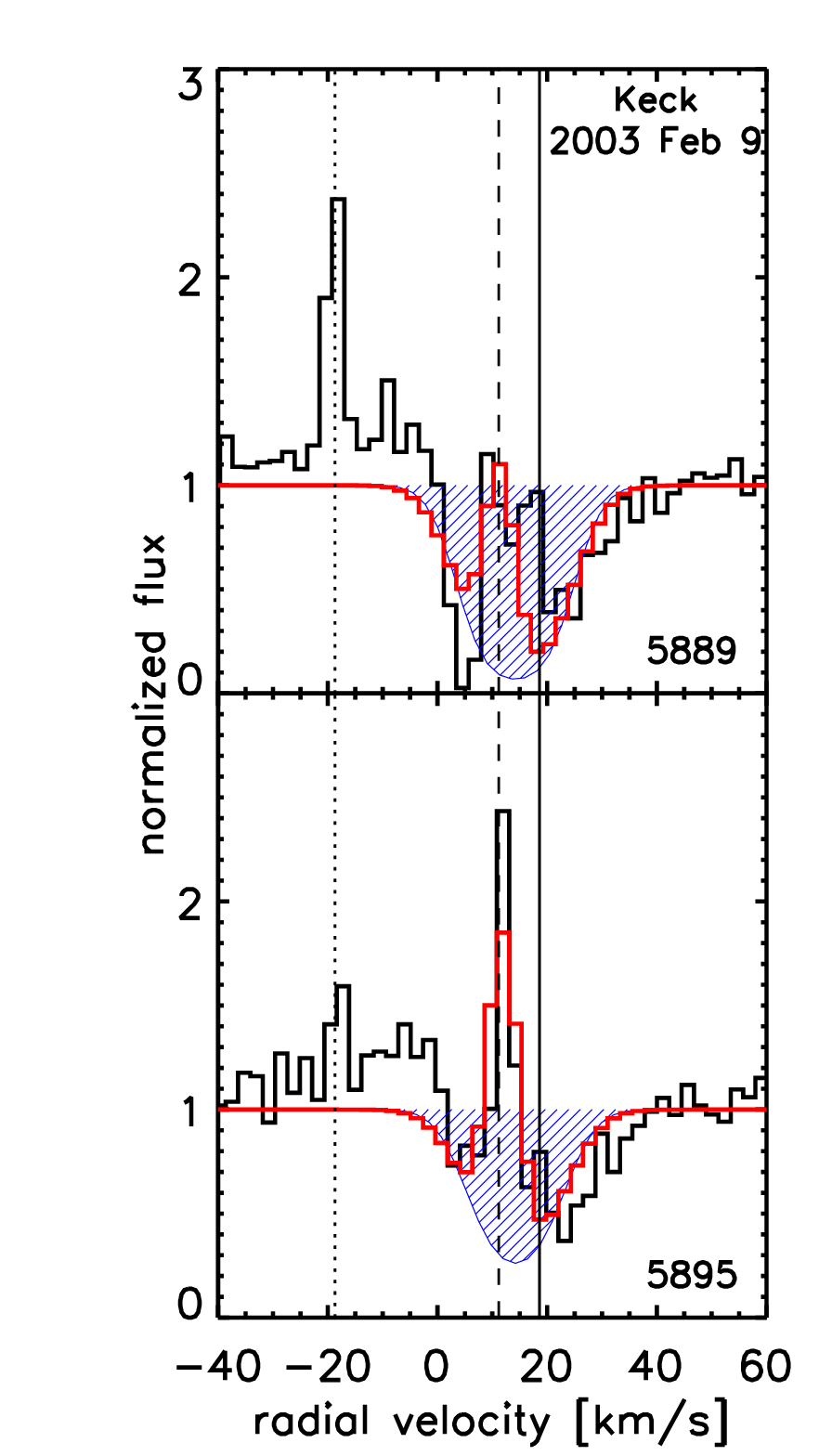}\includegraphics[scale=0.4]{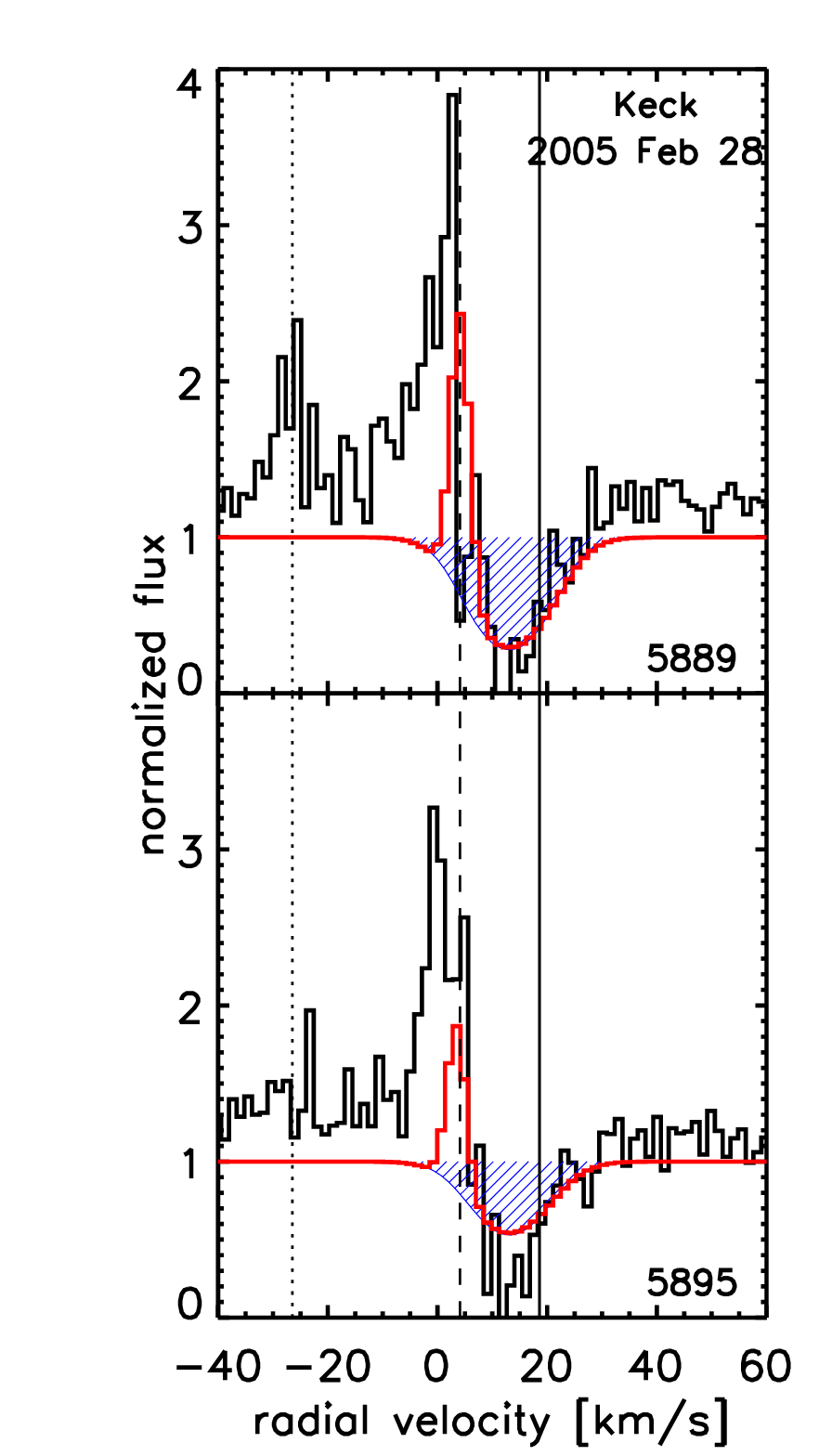}\includegraphics[scale=0.4]{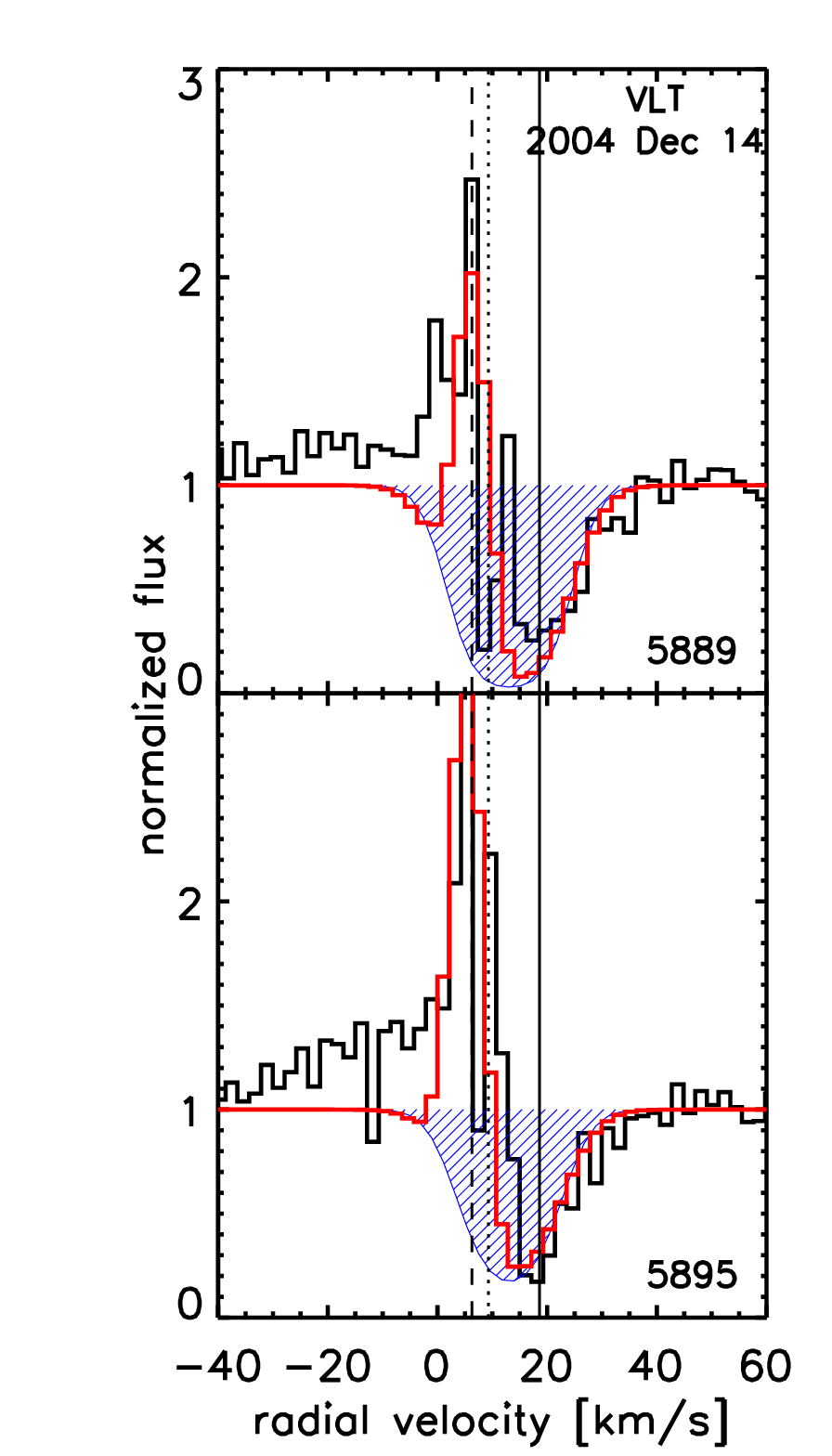} 
\includegraphics[scale=0.4]{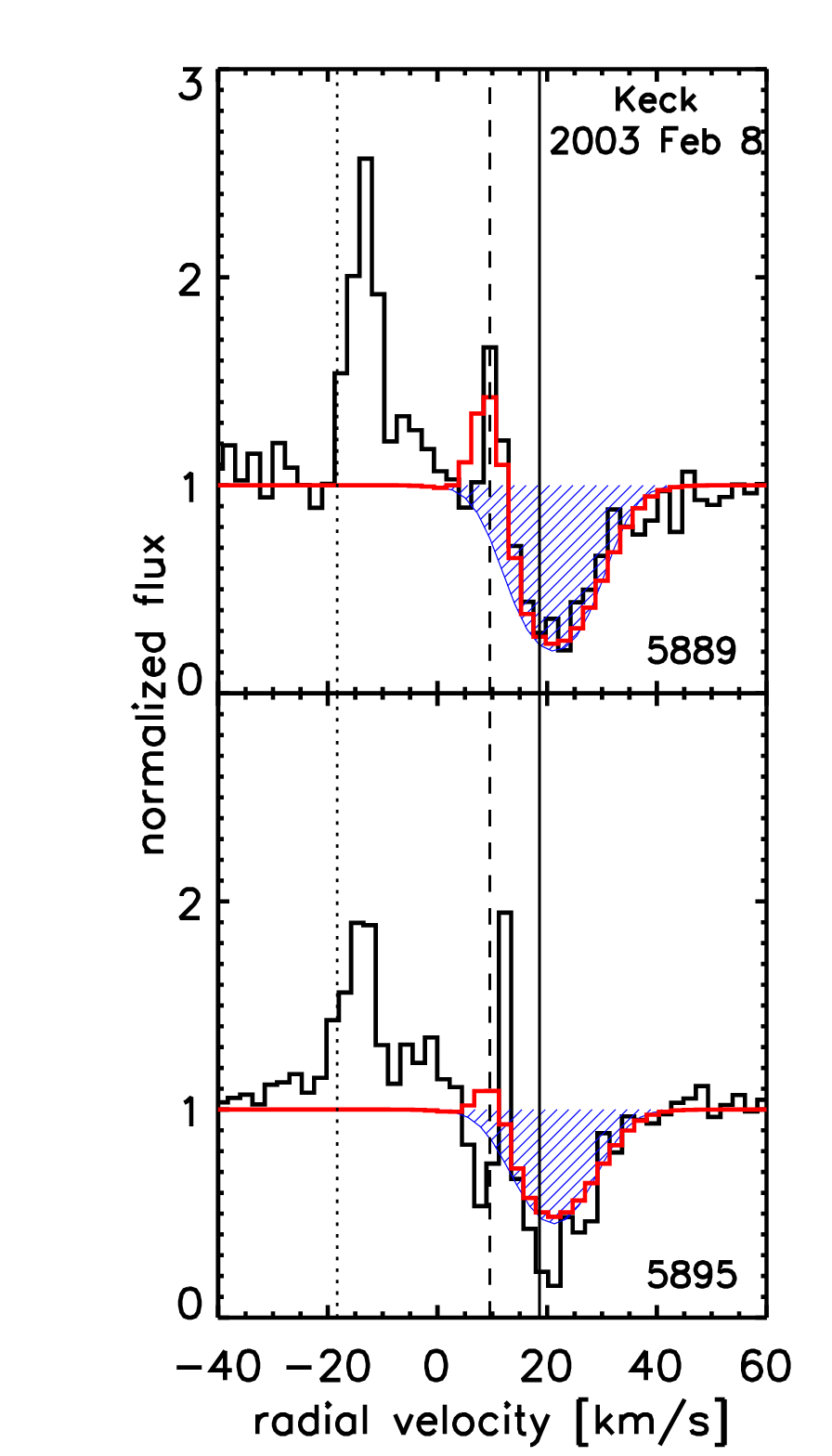}\includegraphics[scale=0.4]{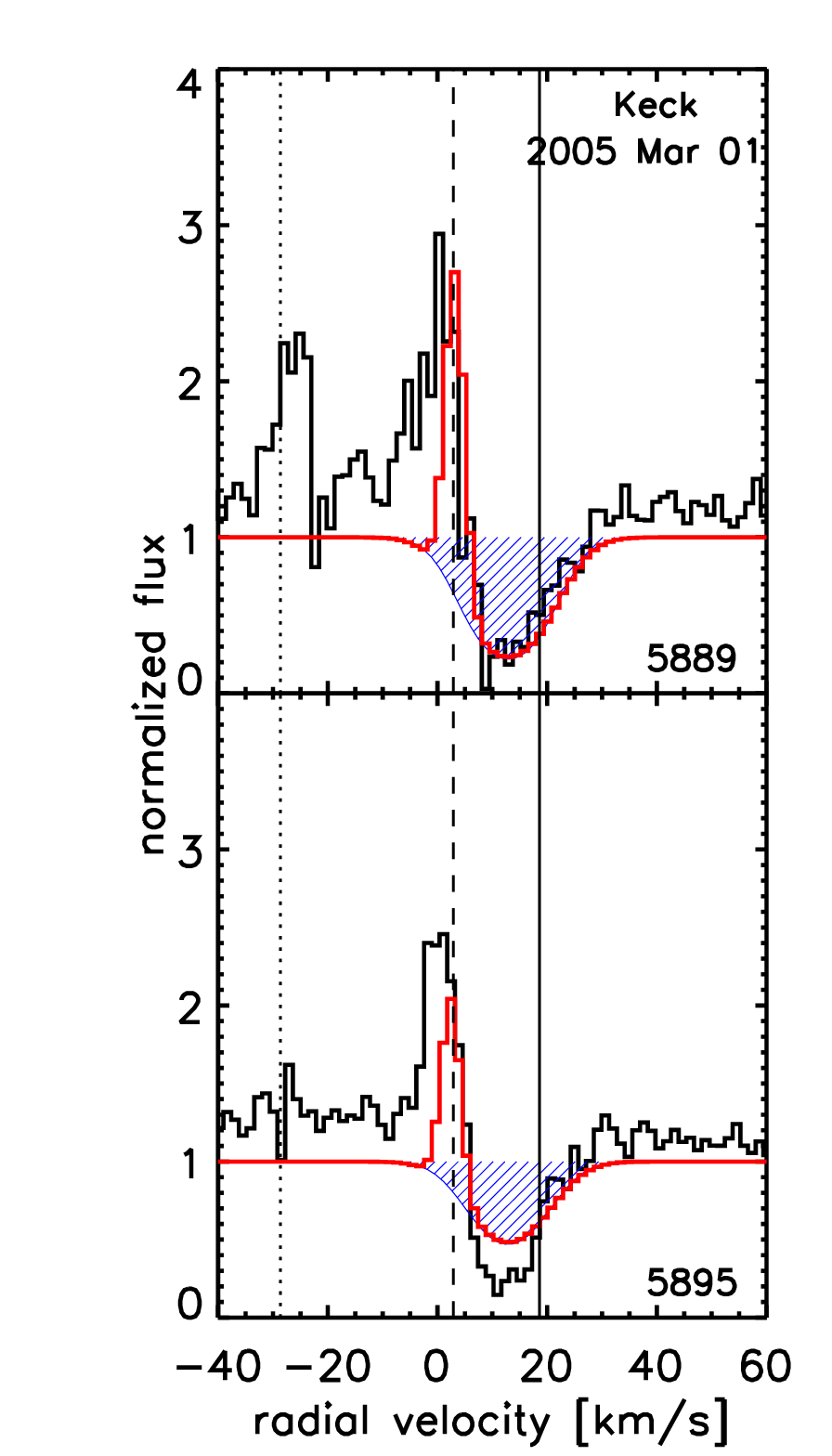}\includegraphics[scale=0.4]{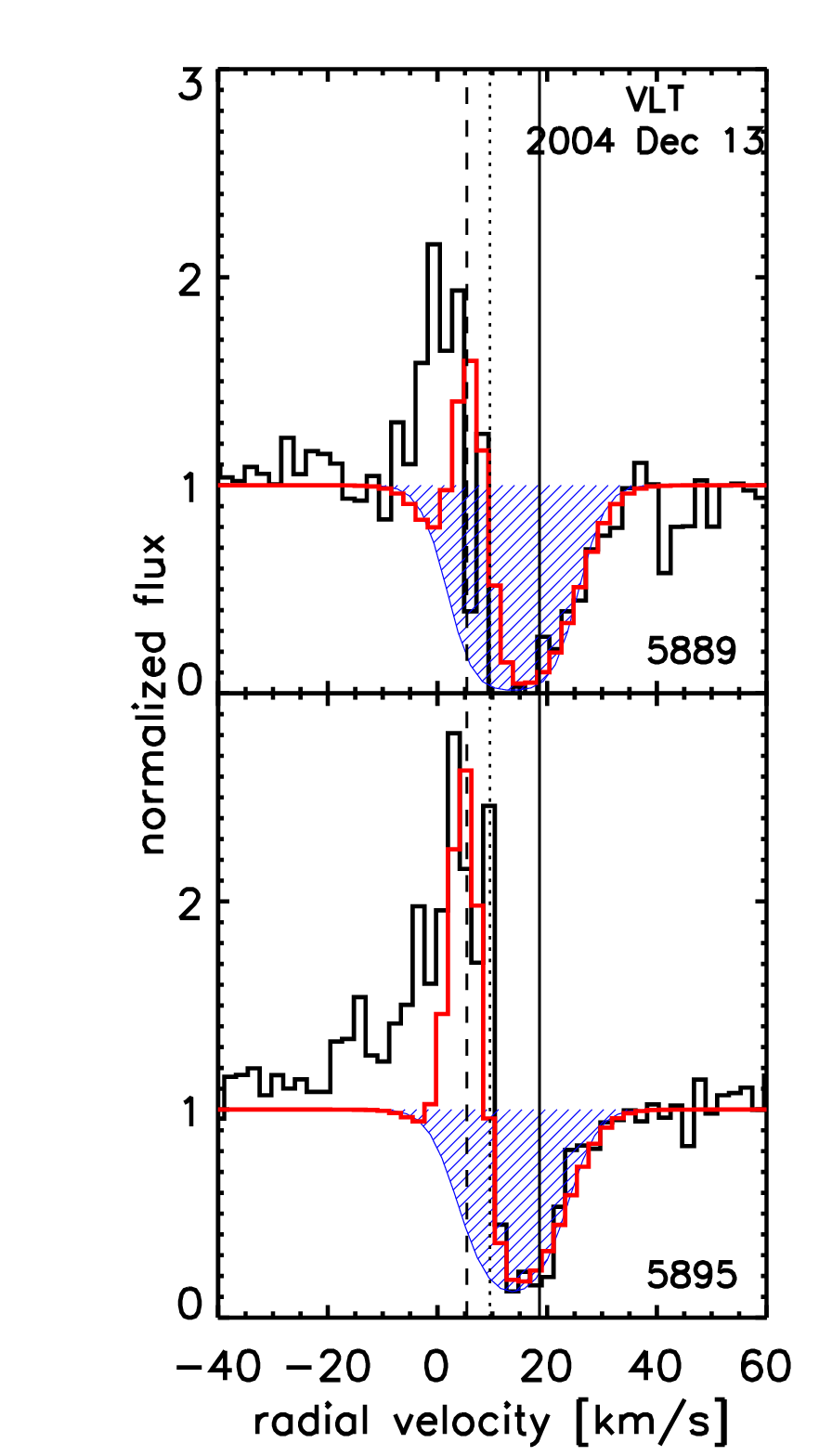}\includegraphics[scale=0.4]{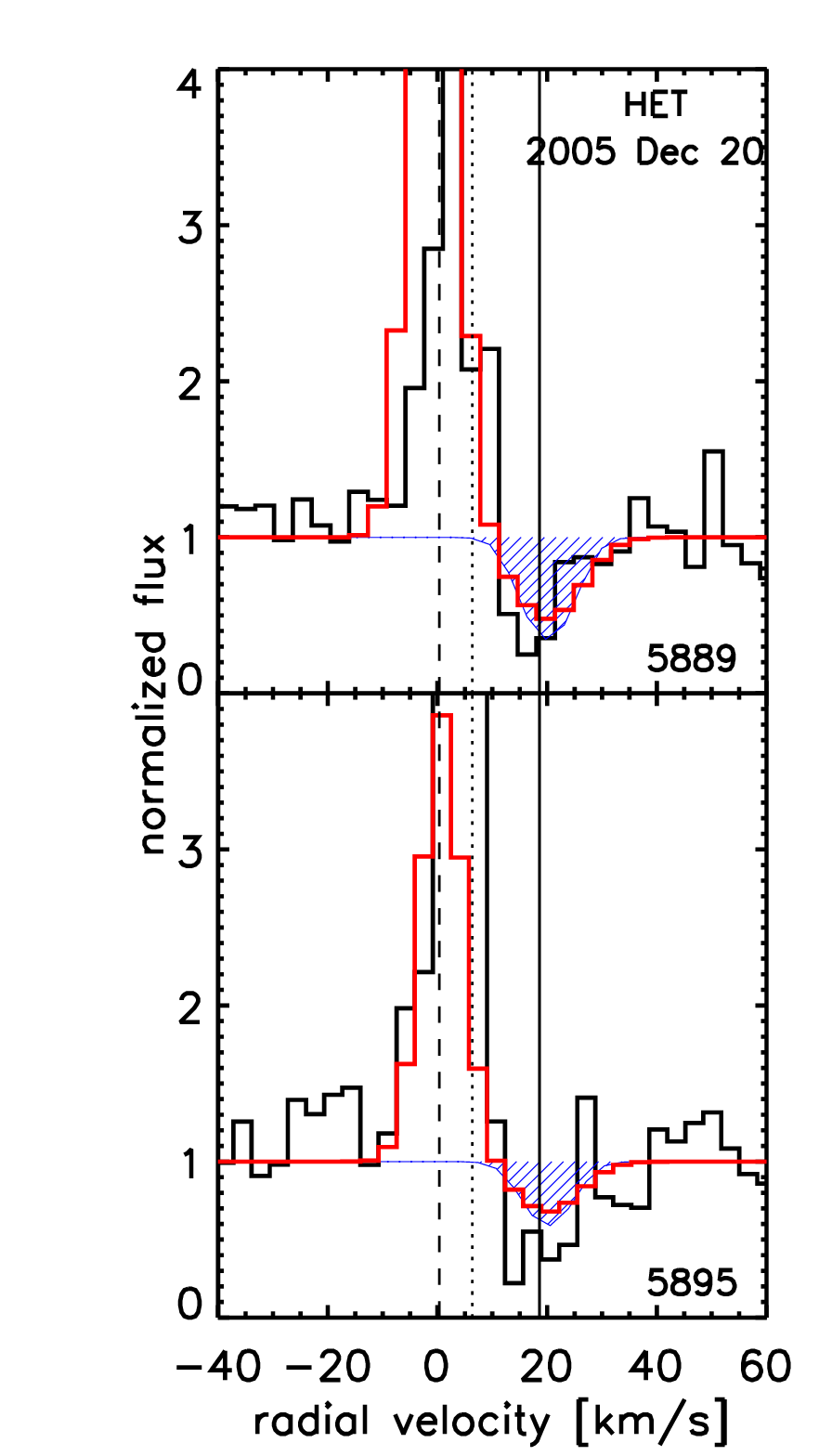} 
\includegraphics[scale=0.4]{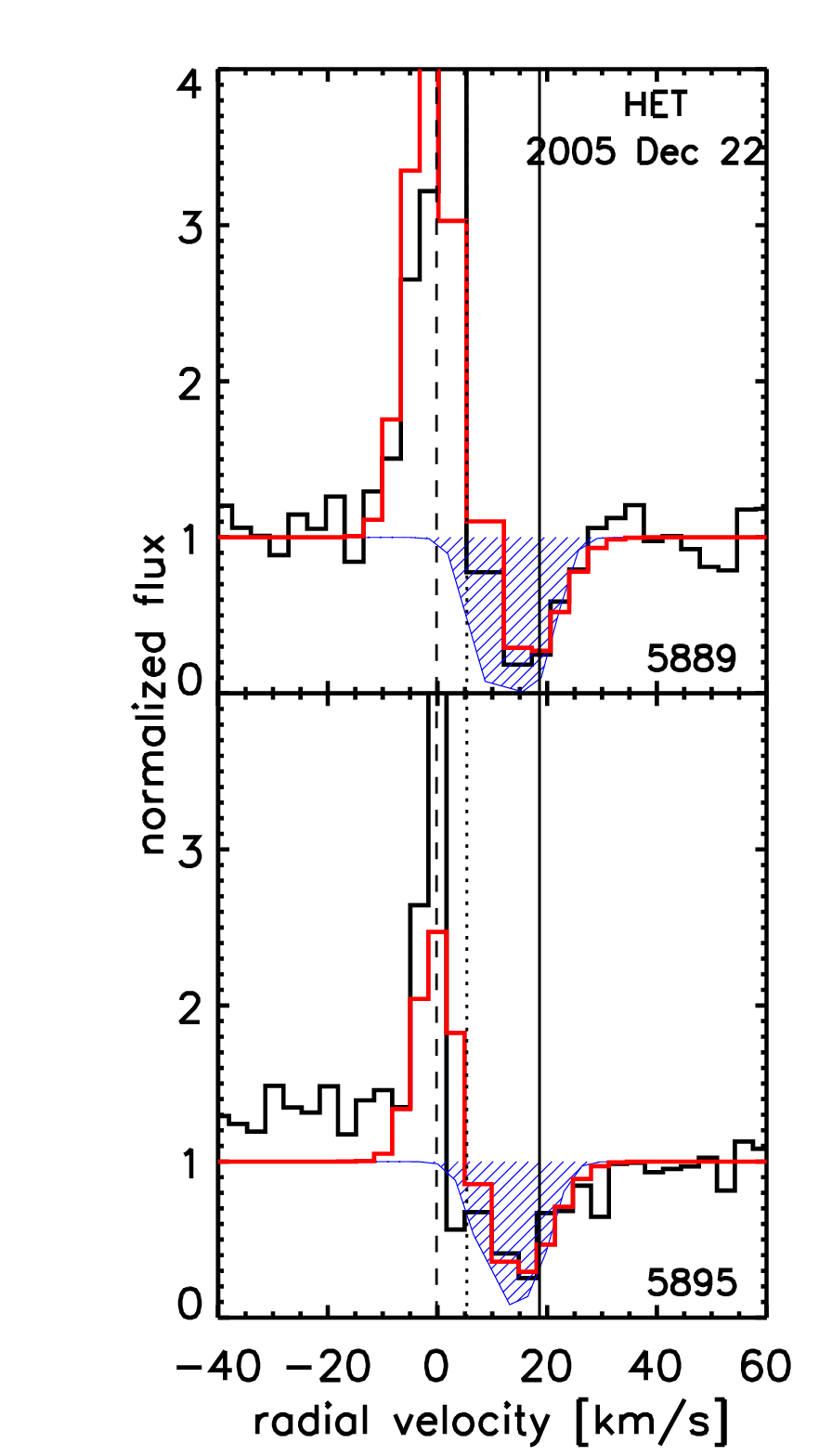}\includegraphics[scale=0.4]{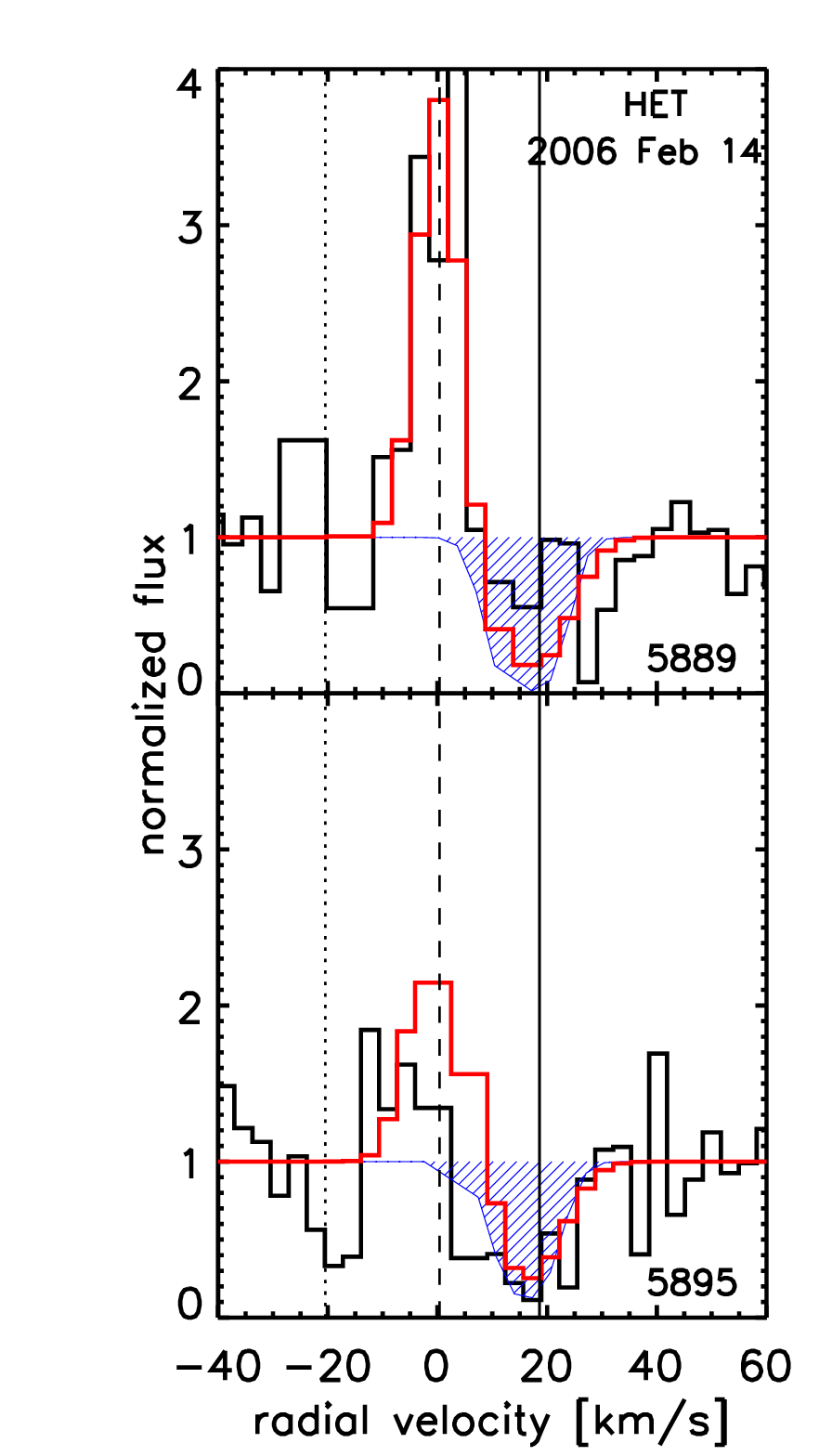}\includegraphics[scale=0.4]{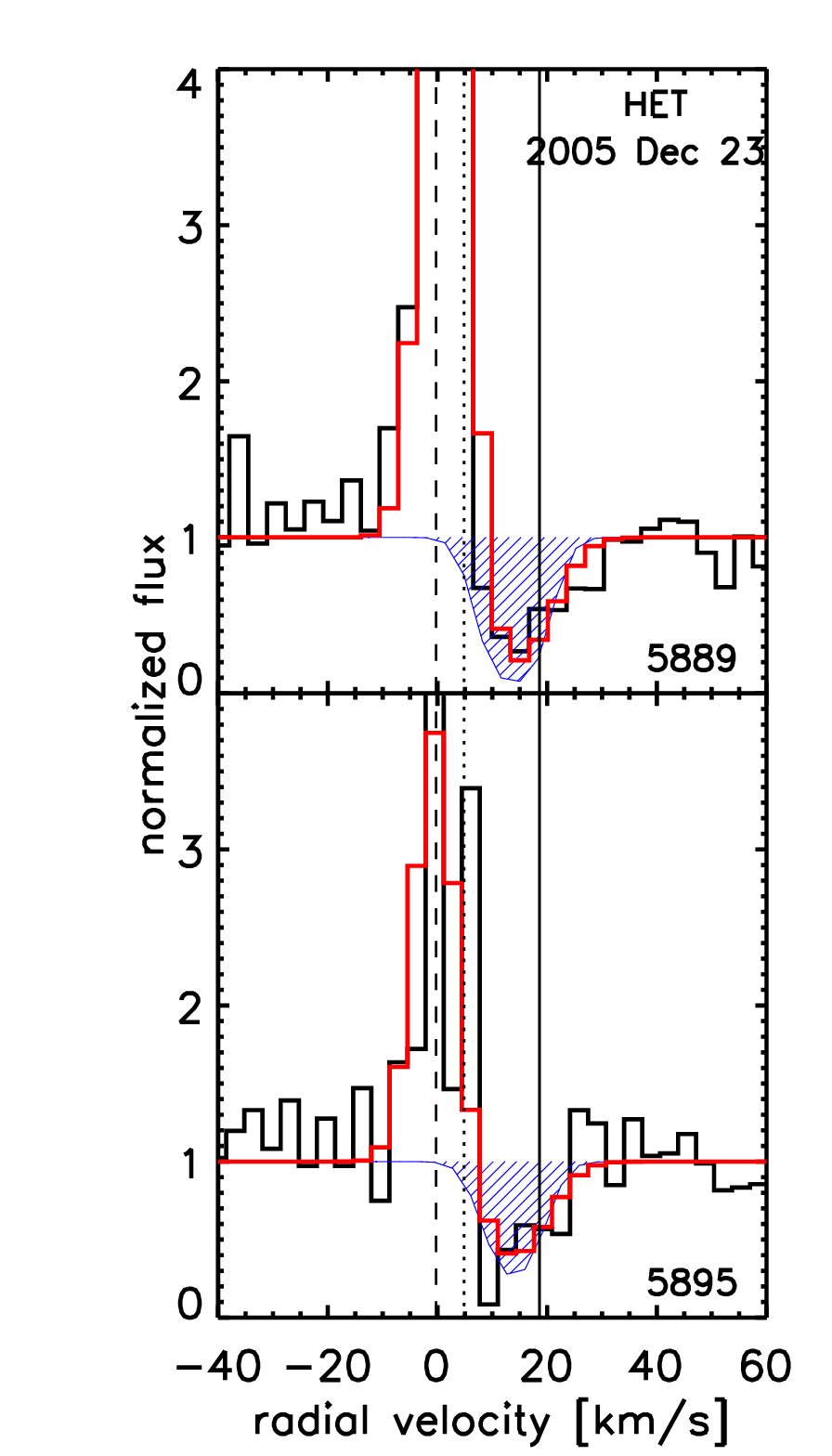}\includegraphics[scale=0.4]{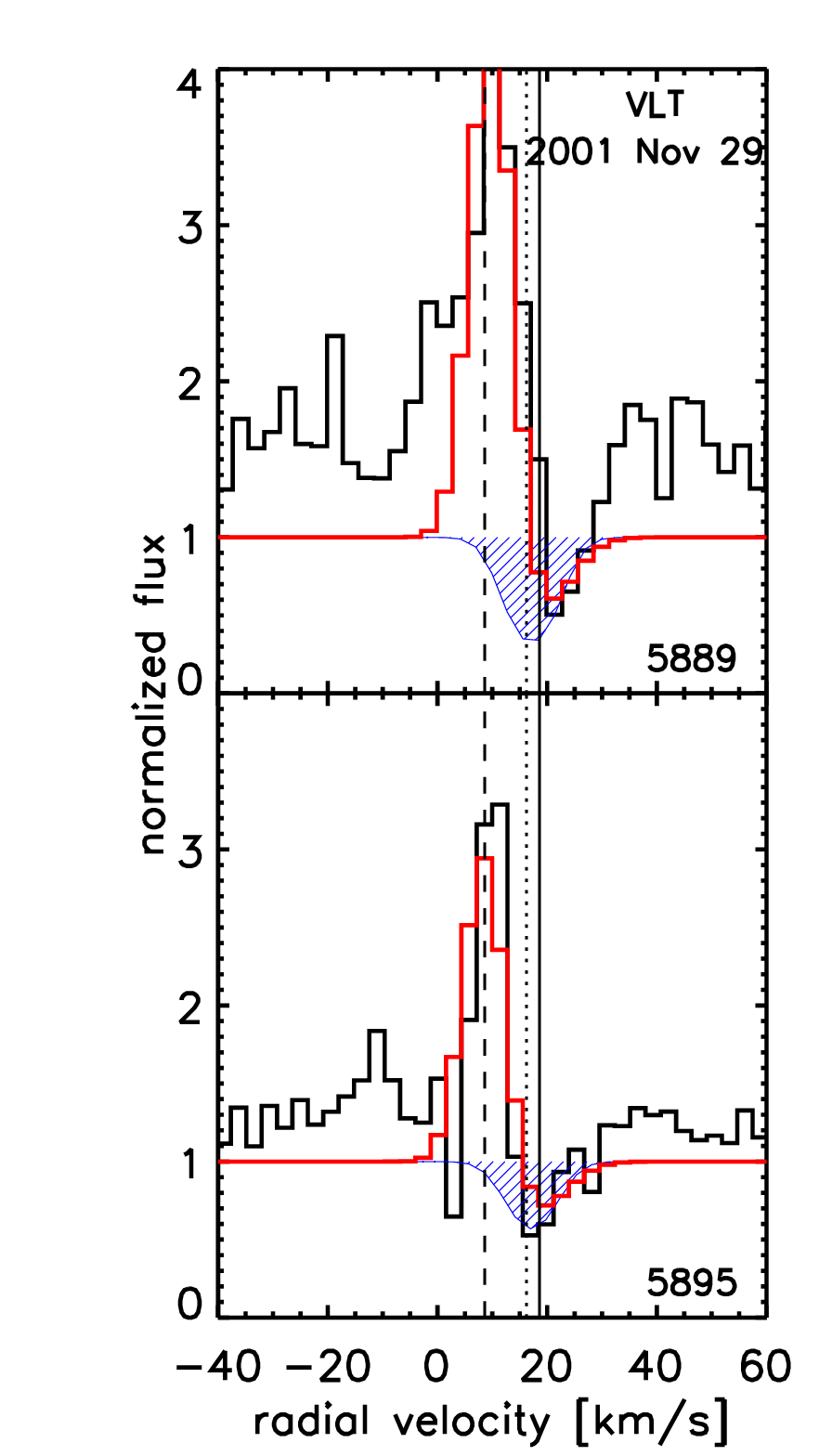} 
\caption{KH~15D spectra from each night ratioed with the comparison star spectrum (black), shown in order of increasing $\Delta$ (see Table~\ref{cds}).  Also shown are the Gaussian absorption feature fits to the excess \ion{Na}{1} absorption (blue filled) and the modeled emission plus absorption (red).  Vertical lines mark the radial velocities of telluric \ion{Na}{1} emission (dotted), Star~A (dashed), and the predicted systemic velocity (solid).}
\label{ratiospec_sm}
\end{figure}

\begin{figure}
\centering
\includegraphics[scale=0.8]{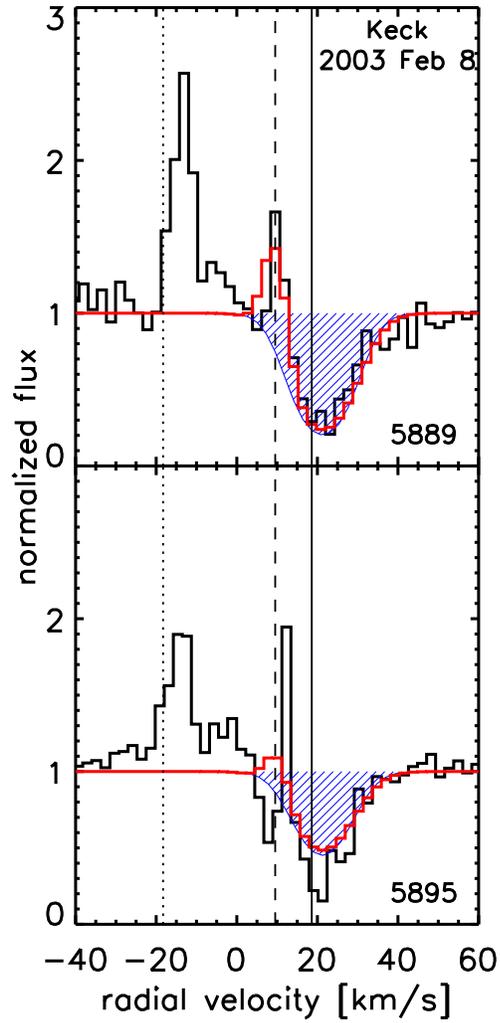} 
\caption{Zoomed in view of the spectrum from Keck taken on 2003 February 8 ratioed with the comparison star spectrum (black), allowing closer inspection of the Gaussian absorption feature (blue filled) and the modeled emission plus absorption (red).  Vertical lines follow Figure~\ref{ratiospec_sm}.}
\label{ratio_big}
\end{figure}

While the dominant feature in these ratioed spectra is the red-shifted excess absorption feature clearly visible in the original spectra, this more detailed look also reveals a more subtle feature.  All of the spectra show evidence of less absorption (or excess emission) at or close to the radial velocity of Star~A. This emission line is especially visible in the 2003 February Keck/HIRES data (see Figure~\ref{ratiospec_sm}) where the radial velocity of Star A is closest to the systemic velocity of 18.6~$\pm$~1.3~km/s \citep{winn06}, placing the emission feature close to the deepest part of the absorption feature in the ratioed spectra.  Regardless of the source of this emission, in order to accurately measure any excess absorption, this emission feature must be taken into account in our model.  

It is possible that the apparent excess emission is due to the spectral mismatch between the comparison star and KH~15D, but we believe that it is likely that Star~A actually does have an emission core within the stellar \ion{Na}{1}~D absorption features.  Three arguments favor the view that the excess emission is real and associated with the star.  We discuss each of these in turn.

First, we note that the emission feature is rather variable. On some nights the emission can be seen very strongly (e.g.\ 2005 December 22) and on others it is barely present (e.g.\ 2003 February 8). While we acknowledge the danger of over-interpreting ratioed spectra, especially deep in the cores of strong lines, the degree of variation does seem larger than can easily be accounted for by simply random noise. If the apparent emission arose only from a mismatch of the spectral types then it should be constant in time.

Another argument is that the emission closely matches the predicted radial velocity of Star~A in each spectrum, even as Star~A's radial velocity varies from -0.27--11.17~km/s during the course of our observations.  This is just what is expected for emission by gas accreting onto the star.  The predicted radial velocities of Star~A are marked in each frame of Figure~\ref{ratiospec_sm} for reference.

Finally, Star~A is known to have an active magnetosphere and bipolar jet \citep{hamilton03}. In this sense it is like a Classical T Tauri Star (CTTS), although the mass accretion rate is probably about an order of magnitude less than for a typical CTTS \citep{mundt09}. Emission in \ion{Na}{1} related to the active magnetospheres and jets has been observed in other T~Tauri stars, particularly when they are oriented such that we view them nearly edge-on to their disks \citep{appenzeller05}, as is the case here. Furthermore, in some spectra the emission lines appear extended to the blue. This is also a common feature of CTTS spectra \citep{appenzeller05} and it may be that we are detecting a rather weak version of the same phenomenon here. 

On the whole, it appears reasonable to suppose that the emission lines seen at the radial velocity of Star~A in all of the spectra in Figure~\ref{ratiospec_sm} are, in fact, arising within the magnetosphere of Star~A, although the physical nature of the excess emission is not critical to the process of correcting for it in the calculation of the absorption column density.

We note that in some spectra (e.g.\ the Keck spectra from 2003 February) there is also a considerable amount of night sky emission which was not fully accounted for during the reduction process. Fortunately, in each case it is sufficiently shifted in radial velocity with respect to KH~15D that we have not needed to remove it from the line profiles because it does not affect the fitting results.

\subsubsection{Quantifying the Excess \ion{Na}{1} Absorption}\label{quant}

In order to most accurately quantify the redshifted excess absorption, we have found it necessary to take into account the variable emission that follows the stellar velocity. Since the \ion{Na}{1}~D line is a doublet we have two absorption and two emission features to model in each spectrum. We have experimented with a variety of different fitting techniques employing constrained and unconstrained Gaussian fits to the emission and absorption components. Given the noise level of the data and uncertainties about the physical source of the features we have, in the end, adopted a procedure quite similar to that employed by \citet{redfield07}. Its implementation here is described below.

For the two \ion{Na}{1}~D lines in the ratioed spectra, we fit a single Gaussian absorption component and a single Gaussian emission component to each.  These four Gaussians (two emission and two absorption) are fit simultaneously, after being convolved with the appropriate line spread function for each instrument.  

The two emission features are fixed at the predicted radial velocity of Star~A for each night of data, with the strength varying freely.  However, the area contained within the Gaussian curve used to model the \ion{Na}{1}~D$_2$ emission line is fixed to be twice that of the Gaussian for the \ion{Na}{1}~D$_1$ line, due to the relative oscillator strengths of the two transitions \citep{morton03}, assuming optically thin \ion{Na}{1} gas.  

The three important parameters for fitting the Gaussian absorption features are the column density, the radial velocity of the maximum absorption, and the Doppler width.  Doppler width, also called the b-parameter or velocity-spread parameter, is the dispersion in velocities multiplied by $\sqrt{2}$.  This measure of the width of the line is believed to be due to the Doppler effect resulting from turbulent and thermal motions \citep{spitzerbook}.  

For the absorption fits, the equivalent widths of the two Gaussian features were also fixed to be 2:1, again corresponding to an assumption of optical thinness.  For our initial fit, the radial velocities and widths of the two Gaussian absorption features were allowed to vary freely and were chosen by $\chi^2$ minimization, with the Doppler widths of the two absorption features fixed to be the same, and the radial velocities fixed to be identical.  This yielded absorption features with an average radial velocity of 16.2~$\pm$~3.3~km/s, consistent with the systemic velocity of 18.6~$\pm$~1.3~km/s \citep{winn06}, and an average Doppler width of 5.6~$\pm$~2.5~km/s.  

For our final absorption measurements, the two Gaussian emission features at the radial velocity of Star A and the two Gaussian absorption features were simultaneously fit to the observed profiles with the Doppler widths fixed to a width that produced the best fit by eye, only varying the column densities (fixed 2:1) and radial velocities (fixed to be identical) of the lines for the $\chi^2$ minimization.  This results in a measurement of the absorbing column density of \ion{Na}{1} necessary to create the observed features.  The results of our fits are listed in Table~\ref{cds} and shown in Figure~\ref{columndensities}.  

For internal consistency in our analysis when comparing spectra from different telescopes and instruments, error bars on the individual spectral datapoints were calculated using the standard deviation of a range of datapoints within the normalized continuum, so that every datapoint within a given spectrum is given the same error value.  The error bars on the column densities result from the standard deviation between three methods of fitting: 1) freely varying all three parameters (radial velocity, Doppler width, and column density), 2) fixing the Doppler widths, and 3) fixing the radial velocity to match the systemic velocity. This is obviously only a crude estimate of the error bars, since they are estimated from only three measured values. However, propagation of errors results in rather similar error bars and the agreement among the various measurements on different nights gives added confidence that these error bars are reasonable. 


One source of error here is the subtraction of telluric \ion{Na}{1}~D emission lines.  In the spectra where the radial velocities of telluric emission lines are closer to the $\sim$18.6~km/s systemic velocity (e.g.\ the 2001 November spectrum), it is more difficult to measure the column densities.  Another source of error is the widely varying strength of the stellar emission lines.  

An additional source of systematic error may be our slightly imperfect comparison star.  Due perhaps to the evolutionary state of KH~15D, as discussed in $\S\ref{measuring}$, an earlier spectral type main sequence star actually matches the spectrum better than an identical spectral type T~Tauri star spectrum, particularly in the deep \ion{Na}{1}~D lines.  However, as noted above, using a non-identical comparison star could introduce a small systematic error in the derived \ion{Na}{1} column densities, as the \ion{Na}{1}~D lines are slightly different in the K5~V star \citep{tripicchio97}.  Such a mismatch is a possible explanation for the apparent excess flux shortward of Star~A's central velocity, mentioned above as also perhaps arising in the magnetospheric emission of Star~A. To a certain extent, our fitting procedure should remove this effect no matter what its cause, but there may well be a residual small effect on the column densities that is systematic in nature.
	
\subsection{Correcting for Interstellar Absorption} \label{ISM}

Figures~\ref{compspec_sm} and~\ref{compspec_big} show that there is indeed excess \ion{Na}{1} absorption present for both of the \ion{Na}{1}~D lines. Additionally, the other absorption lines in the stellar spectrum match up very well with absorption lines in the spectrum of the comparison star.  This alone is strong evidence that there is some absorbing sodium gas toward the KH~15D system.  Whether or not this absorbing material is circumstellar needs further investigation, and in particular, consideration of whether the absorption could arise in the general interstellar medium (ISM) between us and the star.

The only measurement of interstellar sodium in the general direction of KH~15D that we could find in the literature was for S~Mon, a well known spectral type O7 member of NGC~2264 located approximately 25\arcmin\ from KH~15D and likely at about the same distance from Earth. \citet{welty94} measure an absorbing component at roughly the systemic velocity of KH~15D, but much weaker (log~$N_{\rm Na I}$ $\sim$~11.8~cm$^{-2}$), with even weaker absorption extending blueward to a radial velocity of about 0~km/s.  Not only is this feature not strong enough to account for our observations, but the general shape does not match our observations.  However, the distance between the sight line to this star and to KH~15D could result in significantly different interstellar absorption profiles.

To get a better constraint on any possible interstellar medium contribution to the \ion{Na}{1}~D feature in KH~15D, we observed the bright B2~III star HD~47887, a cluster member located only 37\arcsec\ away. A spectrum was obtained with Keck/HIRES on 2003 February 9 to search for evidence of \ion{Na}{1} absorption in that star.  Figure~\ref{bstar} shows the \ion{Na}{1}~D profile of the star and, indeed, two components of interstellar \ion{Na}{1} are clearly visible. We have fit them using the same methods described in $\S\ref{analysis}$ above for KH~15D, shown in the figure. The narrower, deeper component has a column density of log $N_{\rm Na I}$ = 12.002~$\pm$~0.002 cm$^{-2}$, with a Doppler width of 1.35~$\pm$~0.01~km/s and a radial velocity of 6.24~$\pm$~0.05~km/s according to this fit.  The wider, shallower feature has a log $N_{\rm Na I}$ = 12.093~$\pm$~0.001 cm$^{-2}$, with a Doppler width of 10.88~$\pm$~0.02~km/s and a radial velocity of 15.65~$\pm$~0.05~km/s.  

This analysis shows that there may well be a significant contribution to the excess absorption observed in KH~15D from the general interstellar medium. However, it is unlikely that it can fully explain the excess absorption for the following reasons. First, the column densities of \ion{Na}{1} measured in the KH~15D spectra are, in most cases, significantly larger than the total column density measured for HD~47887. Second, only the broad shallow feature is likely to be contaminating our measurements because the narrow deep feature lies at a radial velocity close to that of Star~A in most of our spectra. Its influence would be to reduce the apparent amount of emission in the \ion{Na}{1} emission feature discussed in \S\ref{naemission}, but it cannot be masquerading as the obvious redshifted absorption component seen in Figure~\ref{compspec_sm} because of the radial velocity difference.

The broad, shallow feature, on the other hand, does nearly match the measured radial velocities of the excess absorption in the KH~15D spectra, but is not as deep. It is probably a blend of several weak components at slightly different radial velocities leading to its broad, shallow nature. If this absorption feature appears at exactly the same strength in the direction of KH~15D as it does in HD~47887 only 37$\arcsec$ away, then it is not strong enough to account for all of the absorption, although it does need to be corrected for in assessing how much \ion{Na}{1} gas can reasonably be associated with the KH~15D disk. Our measurements have an average log~$N_{\rm NaI}$ of 12.5~cm$^{-2}$.  Subtracting the contribution of the ISM as measured in the spectrum of HD~47887 leaves log~$N_{\rm NaI} \sim$~12.3~cm$^{-2}$, which is our best estimate of the column density associated with the disk of KH~15D. Formally, the error bar is log~$N_{\rm NaI} \sim$~0.1~cm$^{-2}$, but we note that the possibility of systematic errors and of contributions to the column from gas local to NGC~2264 but not associated with the KH~15D disk means that our measurement may be conservatively viewed as a rather strict upper limit. Certainly the column density of \ion{Na}{1} associated with the putative gaseous disk of KH~15D cannot have a column density much in excess of our measurement of log~$N_{\rm NaI}$ $\sim$~12.3~cm$^{-2}$.


\section{Scale Height of the Gaseous Disk} 

\subsection{Comparing our Measurements to Theory}\label{scaleheight}

\begin{figure}
\centering
\includegraphics[scale=0.5]{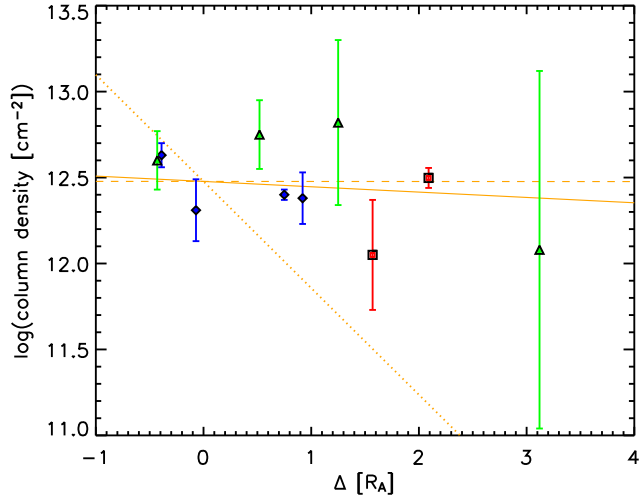} 
\caption{The measured column densities plotted against $\Delta$, the elevation of Star A above the grain disk (Keck: blue diamonds, VLT: green triangles, HET: red squares).  Only two HET/HRS datapoints are shown, because the three datapoints with $\Delta$ close to 2~$R_A$ were averaged together, weighted by their error bars, to one point.  The error bars show the systematic errors, as discussed in $\S\ref{quant}$.  The three orange lines show 3 different scale heights: the scale height of the grain disk ($H$~=~0.7~$R_A$; dotted), the scale height calculated at the inner edge of the gas disk ($H$~=~14~$R_A$; solid), the scale height calculated at the outer edge of the gas disk ($H$~=~2600~$R_A$; dashed).  See $\S\ref{scaleheight}$ for calculations.}
\label{columndensities}
\end{figure}

\begin{figure}
\centering
\includegraphics[scale=0.6]{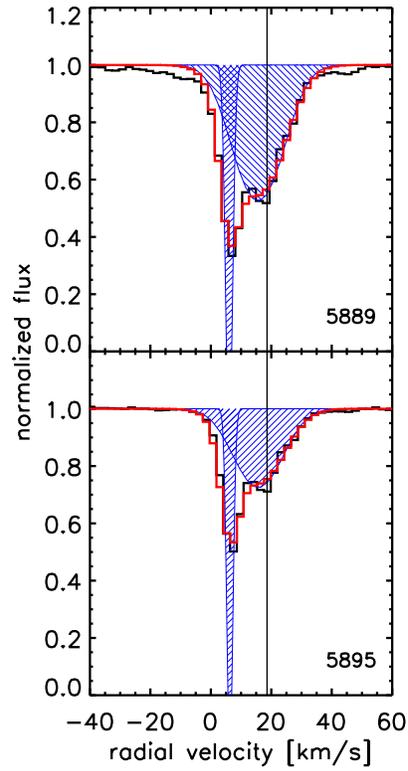} 
\caption{Interstellar \ion{Na}{1} absorption toward the spectral type B2III star HD~47887.  Two absorbing components are seen.  The blue filled areas show the separate Gaussian components before convolution with the line spread function, while the red solid line shows the sum of the two components after convolution.  The solid vertical line shows the systemic velocity of KH~15D for reference.}
\label{bstar}
\end{figure}

Figure~\ref{columndensities} shows that we have not been able to detect a significant variation of \ion{Na}{1} column density as Star~A moved relative to the grain disk edge. This is unfortunate because such a detection would have been clear evidence that the excess \ion{Na}{1} absorption is arising within the gaseous component of the disk. We now inquire whether the failure to detect such variation is an argument that the excess absorption that we have detected could not, in fact, be associated with the disk.

The exact geometry of the KH~15D system is not known and, in particular, we do not know the location of the sharp edge caused by the particulate disk. Our line of sight may intercept the disk (it is perhaps more accurately referred to as a ring in this context) at its inner edge, outer edge, or somewhere between \citep[these possible disk geometries are illustrated very clearly in Figure~6 of][]{winn06}. The inner edge is possibly at the 3:1 resonance with the binary system \citep[$\sim$0.6~AU;][]{herbst08} and the outer edge is probably limited to 5~AU or less in order to precess \citep{chiang04}. Assuming that the particulate disk marks the mid-plane of the gaseous disk and that our line of sight is nearly along the disk plane, we may then equate the apparent distance of Star~A from the opaque disk edge ($\Delta$) with the height of the gas layer that our line of sight to Star~A is probing. Of course, this is a simplification because the disk must be warped and because we are not really looking exactly along it. In the absence of more detailed knowledge of the system's geometry, it is a reasonable approximation.

The scale height ($H$) of the gas layer may be estimated under the assumption of hydrostatic equilibrium in the direction perpendicular to the disk ($z$). Assuming an ideal gas at a temperature ($T$) and constant vertical component of the local gravity ($g_z$), one may write 
$$H =  {g_z^{-1} \biggr[{P \over \rho}\biggr]} = {g_z^{-1} \biggr[{kT \over {\mu m_H}}\biggr]}$$ where $k$ is the Boltzmann constant, $\mu$ is the mean molecular weight of the gas and $m_H$ is the mass of the hydrogen atom. We further assume that the disk gas has a cosmic abundance of helium, is composed of predominantly neutral atoms, and that the Na atoms are well mixed with the H and He. In this case, we may set $\mu = 1.3$.

$T$ is expected to be in the range of about 150 to 400~K, depending on the exact location of the edge. The local vertical component of gravity, $g_z$, may be set by either the self-gravity of the disk or by the vertical component of the stellar gravity. In the first case, $g_z= 2 \pi G \Sigma$, where $\Sigma$ is the mass per unit area in the local disk, and in the second case $g_z = \Omega^2 z$ for small $z$, where  $\Omega$ is the angular velocity of the disk material \citep{goldreich73}. For us, the more interesting limit is the weakest gravitational field that could be present, so we ignore any possible self-gravity. The range of motion of Star~A in our data set is only about 3 stellar radii ($R_A$, where $R_A$~= 1.3~$R_\odot$), so we take $z$~=~1 (in units of $R_A$) for this approximate calculation.

With the assumptions above, we estimate that a putative gas disk in hydrostatic equilibrium would have a measured scale height of $H$~= 14~(2600)~$R_A$ at the inner (outer) edge of the disk, located at 0.6~(5)~AU. These scale heights are shown in Figure~\ref{columndensities}. This is too large a scale height for us to have measured, given the limited range in $\Delta$ probed by Star~A during our observations and the precision of our measurements. Conversely, one can argue that if the the slightly decreasing trend of \ion{Na}{1} column density with $\Delta$, visible in Figure~\ref{columndensities}, is real, then the occulting edge of the gas disk is likely to be the inner edge of the disk and there may well be a contribution to g$_z$ from the self-gravity of the disk (i.e.\ it would need to be a few times larger than the stars alone could provide). 

\subsection{The Settled Grain Disk of KH~15D} \label{graindisk}

We did not detect any statistically significant dependence of the excess \ion{Na}{1} absorption on the height of Star~A above the occulting edge. One expects from the order-of-magnitude analysis in \S\ref{scaleheight} that the scale height will be much greater than the few stellar radii range of $\Delta$ for which we have data. By contrast, the scale height of the grain disk responsible for the extinction must be less than 1~$R_A$. It has been represented in models based on the light curve as a sharp ``knife edge" \citep{herbst02,winn06}, and may be estimated from Figure~1 of the Supplement of \citet{herbst08} as having the value 0.7~$R_A$. While there is uncertainty in all of these estimates it seems clear that the scale height of the particulate disk is at least one order of magnitude less than the scale height of any gaseous disk associated with it.

Furthermore, our lack of detection of a strong \ion{Na}{1} feature associated with the steep increase in extinction at the grain disk boundary demonstrates again how different (evolved) this obscuring matter is from interstellar grains. The extinction in the I~band for our most obscured data point (the first one in Table~\ref{cds}) is 0.9~mag, which for a normal interstellar medium extinction law corresponds to 1.6~mag of extinction in the V~band. The reddening  associated with this amount of extinction would be E(B-V)~$\sim$ 0.5~mag, which would predict a \ion{Na}{1} column density of log~$N_{\rm NaI}$ =~13.2~cm$^{-2}$ \citep{bohlin78}, much higher than the \ion{Na}{1} column density that we actually detect.

Our interpretation of these facts is that the particulate matter in the KH~15D disk has grown large enough to become both optically grey and dynamically separated from the gas layer. This requires particle sizes of a millimeter or so, consistent with the minimum grain size derived by \citet{herbst08} on the basis of a completely different argument. 

It would be interesting to know what the gas-to-dust ratio in the settled particulate layer is. Unfortunately, converting from \ion{Na}{1} column density to total gas column density is difficult for two reasons. First, the ionization fraction of sodium atoms is unknown, and second, it is likely that much of the sodium is depleted onto the solids. If ISM values of the ionization and depletion applied, then there would be significantly more matter in the form of solids than gas. If the gas density is very high, on the other hand, one would expect the solid grains to quickly lose momentum through friction with the gas and spiral into the stars. Unless there were a steady source of new dust, such a structure would disappear on an astronomically brief time scale. It may be, therefore, that the very existence of the particulate disk indicates a fairly low gas to dust ratio. 

\section{Summary and Conclusions} \label{summary}

\ion{Na}{1} lines in the spectrum of the binary WTTS KH~15D have been analyzed in detail. We find an excess absorption component that may be attributed to the gas component of a circumbinary disk. The derived column density is log~$N_{\rm NaI}$ $\sim$~12.5~cm$^{-2}$ with no significant variation associated with the position of Star~A relative to the optically occulting edge.  Subtracting the likely contribution of the ISM leaves log~$N_{\rm NaI} \sim$~12.3~cm$^{-2}$. 

There is no detectable change in the gas column density across the ``knife edge" formed by the edge of the grain disk, indicating that the gas and solids have very different scale heights, with the solids being highly settled. 

If a standard ISM ratio of  $N_{\rm NaI}$ to A$_{\rm V}$ is applied along these lines of sight, there would be a much lower mass of gas than there is mass of solids in this disk. However, the conversion from \ion{Na}{1} column density to total gas column density is complicated by the fact that much of the Na in this disk gas might be either ionized or depleted onto the solids. 

An additional complication is detectable emission in the \ion{Na}{1} feature that follows the radial velocity of Star~A. While this might be due to the slight mismatch in the spectral type of the comparison spectrum, it more likely arises from the active, accreting magnetosphere known to be associated with this (quasi) WTTS.  

Our data support a picture of this circumbinary disk as being composed of a very thin particulate grain layer composed of sand (millimeter) sized objects or larger, which have settled within whatever remaining gas may be present. This phase of disk evolution has been hypothesized to exist as a prelude to the formation of planetesimals through gravitational fragmentation \citep{safronov69, goldreich73} but has not been extensively observed, and is expected to be short-lived if much gas were still present in such a disk.   

Both components of KH~15D are now fully eclipsed by the circumbinary disk for the forseeable future \citep{chiang04}, so our chance to directly measure absorption in this particular disk has passed.  Fortunately, other young, eclipsing, edge-on disk systems such as WL4 \citep{plavchan08} and YLW~16A (Plavchan et~al., in prep.) are being discovered in photometric monitoring surveys, and future spectroscopic studies may allow better understanding of the timescale of dust and gas settling in transition disks.

\acknowledgements{We thank Coryn Bailer-Jones for some of the VLT/UVES data, and Suzan Edwards for the spectrum of LkCa~7.  S.~M.~L.\ would like to thank Roy Kilgard and Peter Plavchan for very helpful advice and discussions regarding this paper.  W.~H.\ acknowledges support by NASA through the Origins of the Solar Systems grant NNX08AK35G.  C.~M.~J.-K.\ acknowledges partial support by NASA through Origins of the Solar Systems grant NNX08AH86G.  J.~N.~W.\ gratefully acknowledges the support of the MIT Class of 1942 Career Development Professorship. J.~A.~J.\ acknowledges support from NSF grant AST-0702821.

Some of the data presented herein were obtained at the W.~M.\ Keck Observatory, which is operated as a scientific partnership among the California Institute of Technology, the University of California and the National Aeronautics and Space Administration. The Observatory was made possible by the generous financial support of the W.~M.\ Keck Foundation.  The authors wish to recognize and acknowledge the very significant cultural role and reverence that the summit of Mauna Kea has always had within the indigenous Hawaiian community.  We are most fortunate to have the opportunity to conduct observations from this mountain.  The Hobby-Eberly Telescope (HET) is a joint project of the University of Texas at Austin, the Pennsylvania State University, Stanford University, Ludwig-Maximilians-Universit{\"a}t M{\"u}nchen, and Georg-August-Universit{\"a}t G{\"o}ttingen.  The HET is named in honor of its principal benefactors, William P.\ Hobby and Robert E.\ Eberly. This paper is based in part on observations collected at the European Organisation for Astronomical Research in the Southern Hemisphere, Chile (Program 074.C-0604A).}

{\it Facilities:} \facility{HET (HRS)}, \facility{Keck:I (HIRES)}, \facility{VLT:Kueyen (UVES)}

\bibliographystyle{apj}


\begin{deluxetable}{lll|cccc}

\tablewidth{0pt} \centering 
\tablecaption{\ion{Na}{1} Column Densities\label{cds}} 															
\tablehead{															
Observation	&	Observation	&	Observatory/	&	$\Delta^{a,b}$	&	Star A RV$^b$	&	BC$^c$	&	log $N_{\rm Na~I}$			\\
Julian Date	&	UT Date	&	Instrument	&	($R_A$)	&	(km/s)	&	(km/s)	&	(cm$^{-2}$)			}
		\startdata													
2453354.7199	&	2004 Dec 15	&	VLT/UVES	&	-0.43	&	7.53	&	8.73	&	12.60	$\pm$	0.17	\\
2452679.8276	&	2003 Feb 9	&	Keck/HIRES	&	-0.39	&	11.17	&	-18.71	&	12.63	$\pm$	0.07	\\
2453429.8250	&	2005 Feb 28	&	Keck/HIRES	&	-0.07	&	4.09	&	-25.18	&	12.31	$\pm$	0.18	\\
2453353.6774	&	2004 Dec 14	&	VLT/UVES	&	0.52	&	6.29	&	9.29	&	12.75	$\pm$	0.20	\\
2452678.8193	&	2003 Feb 8	&	Keck/HIRES	&	0.75	&	9.53	&	-18.29	&	12.40	$\pm$	0.03	\\
2453430.8201	&	2005 Mar 1	&	Keck/HIRES	&	0.92	&	2.89	&	-25.45	&	12.38	$\pm$	0.15	\\
2453352.7976	&	2004 Dec 13	&	VLT/UVES	&	1.25	&	5.35	&	9.56	&	12.82	$\pm$	0.48	\\
2453724.7264	&	2005 Dec 20	&	HET/HRS	&	1.57	&	0.33	&	6.35	&	12.05	$\pm$	0.32	\\
2453726.7222	&	2005 Dec 22	&	HET/HRS	&	2.02	&	-0.18	&	5.35	&	12.73	$\pm$	0.34	\\
2453780.7305	&	2006 Feb 14	&	HET/HRS	&	2.11	&	0.38	&	-20.45	&	12.66	$\pm$	0.96	\\
2453727.7181	&	2005 Dec 23	&	HET/HRS	&	2.16	&	-0.27	&	4.84	&	12.45	$\pm$	0.07	\\
2452242.7446	&	2001 Nov 29	&	VLT/UVES	&	3.12	&	8.60	&	16.23	&	12.08	$\pm$	1.04	\\
\enddata															
\tablenotetext{a}{$\Delta$ is the height of the center of Star~A above the grain disk edge, measured in units of the radius of Star~A ($R_A$ = 1.3 $R_{\odot}$ = 9.0 $\times$ 10$^{10}$ cm).  Note that because $\Delta$ is measured in stellar {\it radii}, none of these spectra were taken when Star~A was completely obscured by the disk.}															
\tablenotetext{b}{Calculated based on a slightly refined version of Model 3 of \citet{winn06}.}															
\tablenotetext{c}{Barycenter correction, which shows the radial velocity telluric \ion{Na}{1} features will have in each of our spectra.}															
\end{deluxetable}											


\end{document}